\documentclass[12pt]{article}
\usepackage{amssymb,amsmath,latexsym}
\usepackage{graphicx}
\usepackage[final]{pdfpages}
\usepackage{verbatim}
\usepackage{color}
\usepackage{amsfonts,amsthm,hyperref,mathrsfs,ulem,tikz}
\usepackage{ulem}
\usepackage{todonotes}
\usepackage{soul}
\usepackage{cancel}

\usetikzlibrary{arrows,patterns}

\title{Analyticity properties of the scattering matrix \\ for matrix Schr\"odinger operators on the discrete line}

\author{Miguel Ballesteros$^1$, Gerardo Franco$^1$, Hermann Schulz-Baldes$^2$
\\
\\
{\small $^1$ IIMAS, UNAM, Cuidad de Mexico, Mexico}
\\
{\small $^2$Department Mathematik, Friedrich-Alexander-Universit\"at Erlangen-N\"urnberg, Germany}
}

\date{ }

\newtheorem{theo}{Theorem}
\newtheorem{defini}[theo]{Definition}
\newtheorem{proposi}[theo]{Proposition}
\newtheorem{lemma}[theo]{Lemma}

\newtheorem{coro}[theo]{Corollary}
\newtheorem{rem}[theo]{Remark}

\newcommand{\CM}{{\mathbb C}}
\newcommand{\NM}{{\mathbb N}}
\newcommand{\RM}{{\mathbb R}}
\newcommand{\SM}{{\mathbb S}}

\newcommand{\ZM}{{\mathbb Z}}

\newcommand{\Aa}{{\cal A}}

\newcommand{\Dd}{{\cal D}}
\newcommand{\Ff}{{\cal F}}
\newcommand{\Gg}{{\cal G}}

\newcommand{\Ss}{{\cal S}}
\newcommand{\Oo}{{\cal O}}
\newcommand{\Tr}{\mbox{\rm Tr}}
\newcommand{\Tt}{{\cal T}}

\newcommand{\Mm}{{\cal M}}
\newcommand{\Cc}{{\cal C}}
\newcommand{\Jj}{{\cal J}}
\newcommand{\Ii}{{\cal I}}

\newcommand{\Kk}{{\cal K}}
\newcommand{\Hh}{{\cal H}}

\newcommand{\Vv}{{\cal V}}
\newcommand{\Ran}{\mbox{\rm Ran}}
\newcommand{\Ker}{\mbox{\rm Ker}}

\newcommand{\one}{{\bf 1}}

\newcommand{\diag}{{\mbox{\rm diag}}}

\newcommand{\overz}{1/z}
\newcommand{\overzb}{1/\overline{z}}
\newcommand{\overzz}{z^{-1}}

\newcommand{\zb}{\overline{z}}

\newcommand{\bsm}{\left(\begin{smallmatrix}} 
\newcommand{\esm}{\end{smallmatrix}\right)}  
\definecolor{GR}{rgb}{.35,.7,.35}

\newcommand{\red}[1]{\textcolor{red}{#1}}

\begin{document}

\maketitle

\begin{abstract}
Explicit formulas for the analytic extensions of the scattering matrix and the time delay of a quasi-one-dimensional discrete Schr\"o\-dinger operator with a potential of finite support are derived. This includes a careful analysis of the band edge singularities and allows to prove a Levinson-type theorem. The main algebraic tool are the plane wave transfer matrices. 
\hfill MSC2010 database: 47A40, 81U05, 47B36

\end{abstract}

\vspace{.5cm}


\section{Overview}
\label{sec-intro}

This paper addresses the stationary scattering theory of self-adjoint matrix Jacobi operators $H$ on $\ell^2(\ZM,\CM^L)$ of the form
\begin{equation}
\label{eq-Hamiltonian}
(H\,u)(n)
\;=\;
u(n+1)
\;+\;V(n)u(n)\;+\;u(n-1)
\;,
\qquad
n\in\ZM
\;,
\end{equation}
where $V(n)=V(n)^*\in\CM^{L\times L}$ is a selfadjoint $L\times L$ matrix and $L\in\NM$ is a fixed number. Such tridiagonal operators are also called Jacobi operators and have a tight connection to orthogonal (matrix) polynomials and via their spectral theory to (matrix-valued) measures on the real line. They are the discrete analogues of Sturm-Liouville operators and many of the techniques such as oscillation theory and Weyl extension theory transpose after suitable modifications. Moreover, these operators are widely used for modeling low-energy phenomena in solid state physics in a one-particle framework. Here the focus will be on an elementary scattering situation where $V(n)$ is non-vanishing only for a finite number of sites $n$, namely the perturbation $V$ of the discrete Laplacian $H_0=H-V$ has finite support. Most standard works \cite{Kur,New,Yaf,Wed,Tes} cover this situation and thus lead to numerous general results. More recently, the case of non-linear  potentials has also been addressed, see \cite{CN2018,N2018} and references therein. Furthermore, the scalar case $L=1$ has been treated in detail by Hinton, Klaus and Shaw a long time ago \cite{HKS}, see also the recent contribution \cite{IT}. For continuous matrix-valued Schr\"odinger operators there are works by Klaus \cite{Kla} and Aktosun, Klaus and Van Der Mee \cite{AKV} as well as a more recent contributions by Aktosun, Klaus and Weder \cite{AW,AKW,AW2}. Also the half-space version of the above discrete matrix Schr\"odinger equation has been analysed \cite{ACP,NRT}. Inverse problems have been addressed in numerous works (see for example \cite{AM1}, \cite{RW2, AW2,AVP, CK,DT,IK}, and the references therein). In spite of all these prior works, this paper has a several novel facts and features that solidify our understanding of stationary scattering theory:
\begin{itemize}
\item The analytic structure of the scattering matrix and the time delay in complex energy $E$ or rather the parameter $z$ defined by $E=z+z^{-1}$ is studied in detail.
\item The unitarity relation of the scattering matrix is extended to complex energies, including the band edge thresholds. 
\item The analyticity, $\Jj$-unitarity and multiplicativity of the newly introduced plane wave transfer matrices are consistently used in the arguments, leading to elementary algebraic proofs of the main results.
\item The connections between the plane wave transfer matrices and the standard transfer matrices used in the theory of Jacobi operators is established.
\item As an application, a Levinson-like theorem connecting the total time delay to the number of bound and half-bound states is proved using purely complex analytic means (namely, the argument principle). Strictly speaking this is a new result, even though there are numerous prior contributions (in particular, \cite{AW}).
\end{itemize}

\vspace{.2cm}

Let us begin by recalling the elementary spectral analysis of $H$. By Weyl's theorem, the essential spectrum of $H$ is given by the (absolutely continuous) spectrum $\sigma(H_0)=[-2,2]$ of $H_0$. It will be part of the results below (Theorem~\ref{theo-Levinson}) that the remainder of the spectrum of $H$ only consists of a finite number $J_b$ of eigenvalues  $E_1,\ldots,E_{J_b}\in\RM$  (listed with their multiplicity) outside of $[-2,2]$. The associated eigenvectors are called bound states and make out the full point spectrum $\sigma_p(H)=\{E_1,\ldots,E_{J_b}\}$ of $H$, notably there are no embedded eigenvalues in the present situation. At the band edge thresholds $E=\pm 2$, there may be a number $J_h$ of further so-called half-bound states that will be discussed as well.

\vspace{.2cm}

For the analysis of the scattering problem and the analytic continuation of the scattering matrix, it is convenient to study the Jacobi equation $Hu=Eu$ for complex energies $E$ and matrix-valued or vector-valued functions of the form $u:\ZM\to\CM^{L\times L}$ or $u:\ZM\to\CM^{L}$ where $H$ is given by the same prescription as in \eqref{eq-Hamiltonian}. The  Jost solutions are particular such solutions which are fixed by their asymptotic behavior at $\pm\infty$ dictated by solutions of the free Jacobi equation $H_0u=Eu$ associated to the free Hamiltonian $H_0$. For any $E\in\CM\setminus\{-2,2\}$ there are two such free solutions 
$$
u_0^z(n)
\;=\;
z^n\,\one
\;,
\qquad
u_0^{\overz}(n)
\;=\;
z^{-n}\,\one
\;,
$$
where $\one\in\CM^{L\times L}$ the identity matrix and $z,z^{-1}\in\CM$ are the two solutions of
\begin{equation}
\label{eq-EnergyZ}
E
\;=\;
z\,+\,z^{-1}\;.
\end{equation}
Note that the map $z\in\CM\setminus\SM^1\mapsto E\in\setminus[-2,2]$ is two-to-one and the associated Riemann surface for $E$ consists of two sheet that are connected through branching on $[-2,2]$. Now $(u_0^z,u_0^{\overz})$ is a fundamental solution of $H_0u=Eu$. Let us point out that for $E\not\in(-2,2)$, these functions $u_0^z$ are exponentially increasing or decreasing in $n$, while for $z\in\SM^1\setminus\{-1,1\}$ they are the well-known plane wave solutions. Further note that for $z=\pm 1$ corresponding to the band edges $E=\pm 2$, there is only one free solution $u^{\pm 1}_0$. There is, however, again a second solution $v_0^{\pm 1}$ of $H_0u=\pm 2u$ given by $v_0^{\pm 1}(n)=(\pm 1)^n\,n$. This second solution is hence increasing linearly in $n$. Now the Jost solutions are defined.


\begin{defini}
\label{def-Jost}
The Jost solutions $u^z_\pm:\ZM\to\CM^{L\times L}$ of $Hu=Eu$ with $E=z+z^{-1}$ are defined by the equalities  
$$
u^z_+(n)
\;=\;
u^z_0(n)
\;,
\qquad
u^z_-(-n)
\;=\;
u^z_0(-n)
\;,
$$ 
holding for large enough $n\in \mathbb{N}$ such that the support of $V$ is contained in $(-n,n)$. For $|z|<1$, the Jost solution $u_+^z$ is called right-decreasing and $u_-^z$ left-increasing, while $u_+^{\overz}$ is called right-increasing and $u_-^{\overz}$ left-decreasing.
\end{defini}

There are in total $2L$ linearly independent solutions $u:\ZM\to\CM^L$ of the equation $Hu=Eu$. For $E\in\CM\setminus \{-2,2\}$, namely $z\in\CM\setminus\{-1,0,1\}$, the columns of the matrix $(u_+^z, u_+^{\overz})$ provide a basis of the space of solutions. The same holds for and $(u_-^z,u_-^{\overz})$. One can thus expand any solution w.r.t. one of these basis. This justifies that the following definition is possible.

\begin{defini}
\label{def-Mtransfer}
For $z\in\CM\setminus\{-1,0,1\}$, the plane wave transfer matrix $\Mm^z\in\CM^{2L\times 2L}$ is defined by
\begin{equation}
\label{eq-MMatDef}
\big(
u^z_- ,u^{\overz}_-
\big)
\;=\;
\big(
u^{z}_+ , u^{\overz}_+
\big)  
\,\Mm^z
\;.
\end{equation}
The $L\times L$ entries of the transfer matrix are denoted by
\begin{equation}
\label{eq-MDef3}
\Mm^z
\;=\;
\begin{pmatrix}
M^{\overz}_- & N^{{z}}_- \\
N^{\overz}_- & M^z_-
\end{pmatrix}
\;.
\end{equation}
\end{defini}

Note that \eqref{eq-MMatDef} contains the equations $u_-^z=u_+^zM^{\overz}_-+u_+^{\overz}N^{\overz}_-$ and $u_-^{\overz}=u_+^zN^{z}_-+u_+^{\overz}M^{z}_-$, which are actually identical if $z$ is replaced by $z^{-1}$. Provided that $M^{z}_-$ is invertible, one can rewrite the second equation as $u_+^{\overz}=u_-^{\overz}(M^{z}_-)^{-1}- u_+^zN^{z}_-(M^{z}_-)^{-1}$. In a similar manner, one can expand ${u^z_+}$ in terms of  ${u_+^{\overz}}$ and ${u_-^z}$, provided that $M^{\overz}_-$ is also invertible. The scattering matrix is then defined by a relation similar to \eqref{eq-MMatDef}. To shorten notations in the following, let us introduce the set $\CM_0$ of those points in $z\in \CM\setminus\{-1,0,1\}$ at which both $M^{\overz}_-$ and $M^z_-$ are invertible. It will be shown (Corollary~\ref{coro-C0} below) that all points in the closed unit disc are in $\CM_0$, except those $z\in (-1,1)$ for which $z+z^{-1}$ is an eigenvalue of $H$.

\begin{defini}
\label{def-Scat}
For any $z\in\CM_0$, the scattering matrix $\Ss^z\in\CM^{2L\times 2L}$ expresses the Jost solutions $u^z_-$ and $u^{\overz}_+$ in terms of $u^{z}_+$ and $ u^{\overz}_-$, namely it is defined by 
\begin{equation}
\label{eq-SMatDef}
\big(
u^z_- , u^{\overz}_+
\big)
\;=\;
\big(
u^{z}_+ , u^{\overz}_-
\big) 
\,\Ss^z
\;.
\end{equation}
The $L\times L$ entries of the scattering matrix are denoted  by
$$
\Ss^z
\;=\;
\begin{pmatrix}
T_+^z & R^z_- \\ R^z_+ & T^z_- 
\end{pmatrix}
\;,
$$
and are called the transmission coefficients $T^z_\pm$ and reflection coefficients $R^z_\pm$.
\end{defini}

\begin{figure}
\centering
\includegraphics[width=9cm]{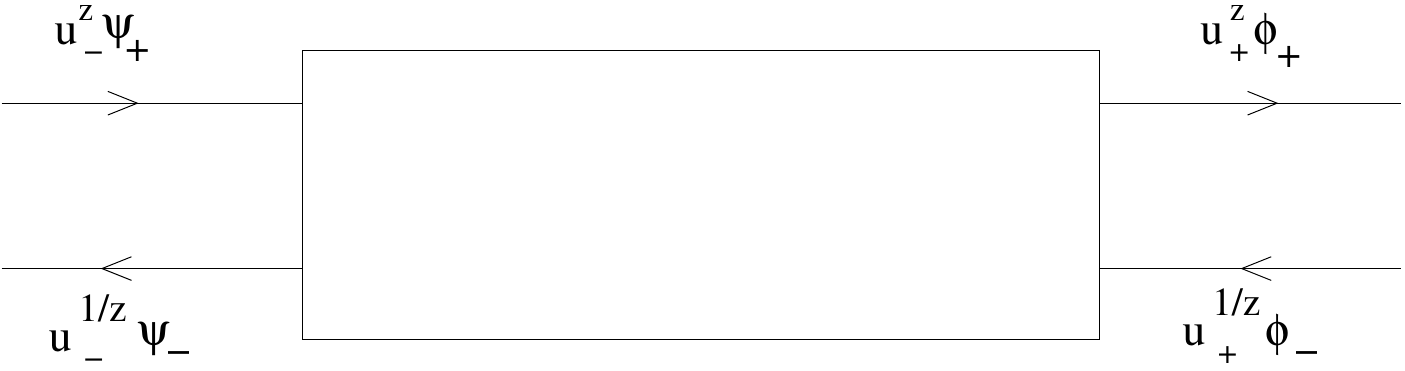} 
\caption{Schematic representation of the scattering process for $|z|<1$. The arrows indicate the direction in which the solutions decrease.
}
\label{fig-Schematic}
\end{figure}

Of course, there are tight connections between the coefficient matrices of $\Mm^z$ and $\Ss^z$. For example, the relation $u_+^{\overz}= u_-^{\overz}(M^{z}_-)^{-1}-u_+^zN^{z}_-(M^{z}_-)^{-1}$ noted above directly implies that $T^z_-=(M^{z}_-)^{-1}$ and $R^z_-=-N^{z}_-(M^{z}_-)^{-1}$. Let us next provide a list of basic algebraic properties of the plane wave transfer matrix and scattering matrix. Some will be expressed in terms of the real Pauli matrices of size $2L\times 2L$ with $L\times L$ blocks:
\begin{equation}
\label{eq-Pauli}
\Kk
\;=\;
\begin{pmatrix}
0 & \one \\ \one & 0 \end{pmatrix}
\;,
\qquad
\Ii
\;=\;
\begin{pmatrix}
0 & -\one \\ \one & 0
\end{pmatrix}
\;,
\qquad
\Jj\;=\;
\begin{pmatrix}\one & 0 \\ 0 & -\one \end{pmatrix}
\;.
\end{equation}
Furthermore, let $\Mm^*$ denote the adjoint (conjugate transpose) of a matrix $\Mm$ and $\imath=\sqrt{-1}$.

\begin{proposi}
\label{prop-TransScatProp}

\begin{itemize}

\item[{\rm (i)}] The plane wave transfer matrix satisfies for all $z\in\CM\setminus\{-1,0,1\}$
\begin{equation}
\label{eq-MJunitarityIntro}
(\Mm^{\overzb})^*\,\Jj\,\Mm^z
\;=\;
\Jj\;,
\qquad
\Mm^{z}\,\Jj\,(\Mm^{\overzb})^*
\;=\;
\Jj\;.
\end{equation}
For $z\in\SM^1\setminus\{-1,1\}$, this shows that  $\Mm^z$ is $\Jj$-unitarity, namely $(\Mm^z)^*\Jj\Mm^z=\Jj$. 

\item[{\rm (ii)}] For $z\in\RM$, the plane wave transfer matrix is $\Ii$-unitary, namely $(\Mm^z)^*\Ii\Mm^z=\Ii$.

\item[{\rm (iii)}]
For all $z\in\CM_0$, one has
$$
(\Ss^{\overzb})^*\,\Ss^z
\;=\;
\one\;,
\qquad
\Ss^{\zb}\;=\;\Kk\,(\Ss^z)^*\,\Kk
\;.
$$
For $z\in\SM^1\setminus\{-1,1\}$, this implies that $\Ss^z$ is unitary, $(T_+^{\zb})^*=T_-^z$ and $(R_\pm^{\zb})^*=R_\pm^z$. 

\item[{\rm (iv)}]
If $V=0$, one has $\Mm^z=\one$ and $\Ss^z=\one$ for all $z\in\CM_0$.

\item[{\rm (v)}]
Both $\Mm^z$ and $\Ss^z$ are meromorphic on $\CM$. 
\end{itemize}
\end{proposi}

The proofs of (i) and (ii) are given Section~\ref{sec-PlaneWaveRep}, and that of (iii) and (v) in Section \ref{sec-Scat}.  Item (iv) is obvious. A detailed analysis of the singularities and zeros of $\Mm^z$ and $\Ss^z$ and their matrix entries make up a large part of the paper.
Before coming to the main results of the paper, let us make several remarks on structural properties of the plane wave transfer matrices and the scattering matrix.

\begin{rem}
{\rm
There are two main reasons justifying the terminology {\it plane wave transfer matrix} for $\Mm^z$. First of all, it reflects that the plane wave solutions from the right are transferred to the plane wave solutions on the left, see Figure~1. The second reason is that $\Mm^z$ is the ordered product of single site plane wave transfer matrices defined by
\begin{equation}
\label{eq-TransferMExplicit0}
{\Mm}^z(n)
\;=\;
\begin{pmatrix}
\one & 0 \\ 0 & \one 
\end{pmatrix}
\;+\;
\imath \nu^z\,
\begin{pmatrix}
V(n) & z^{-2n}V(n) \\ -z^{2n} V(n) & -V(n) 
\end{pmatrix}\,
\;,
\end{equation}
where 
\begin{equation}
\label{eq-NuDef}
\nu^z\;=\;\frac{\imath}{z-\overzz}
\end{equation}
is a quantity that will be appear in many formulas below. It satisfies $\nu^z=-\nu^{\overz}=(\nu^{\overzb})^*$, is analytic in $\mathbb{C}\setminus\{-1,0,1\}$ and has poles of first order at $\pm 1$.  Then the multiplicativity, a defining property of any type of transfer matrices, is given by
$$
\Mm^z
\;=\;
\Mm^z(K_+)\Mm^z(K_+-1)\cdots\Mm^z(K_-+1)
\;,
$$
where $[K_-+1,K_+]\cap\ZM$ contains the support of $V$. The multiplicativity will be explained in detail in Section~\ref{sec-PlaneWaveRep} and exploited later on. As far as we know, one novel point of this paper is how to obtain $\Mm^z$ and $\Mm^z(n)$ from the standard transfer matrix used in the theory of (block) Jacobi matrices, see Section~\ref{sec-PlaneWaveRep}.
}
\hfill $\diamond$
\end{rem}

\begin{rem}
{\rm
In this remark, $z\in\SM^1$. Then there are two perspectives on the passage from $\Mm^z$ to $\Ss^z$. The first one is via the singular value decomposition and is standard in the physics literature. It goes back at least to \cite{MPK}, and is, from a mathematical perspective, also developed in the appendix of \cite{Sad} so that no further details are given here. The second one is based on the fact that the graph of $\Jj$-unitary matrix is Lagrangian w.r.t. $\Jj\oplus\Jj$ and is now described in some detail. Let us first recall (in slightly different variant than in \cite{SB2,SB,MSB})  that to any $\Jj$-unitary $\Mm$ (namely, satisfying $\Mm^*\Jj\Mm=\Jj$ so that also $\Mm\Jj\Mm^*=\Jj$), there is an associated unitary $\Vv(\Mm)\in\CM^{2L\times 2L}$ by
\begin{equation}
\label{eq-VM}
\Vv(\Mm)
\;=\;
\begin{pmatrix}
(A^*)^{-1} & -BD^{-1} \\
D^{-1}C & D^{-1}
\end{pmatrix}
\;=\;
\begin{pmatrix}
(A^*)^{-1} & -BD^{-1} \\
B^*(A^*)^{-1} & D^{-1}
\end{pmatrix}
\;,
\qquad
\Mm\;=\;
\begin{pmatrix} A & B \\ C & D
\end{pmatrix}
\;.
\end{equation}
This unitary describes geometrically the twisted graph of $\Mm$, see \cite{SB2}. It satisfies 
$$
\Vv(\Mm)\;=\;\Vv(\Mm^{-1})^*\;=\;\Jj \,\Vv(\Mm^*)^*\Jj
\;,
$$ 
and, moreover, there is a tight connection between the spectral theory of $\Mm$ and $\Vv(\Mm)$, namely the multiplicity of $1$ as eigenvalue of $\Mm$ is equal to the multiplicity of $1$ as eigenvalue of $\Vv(\Mm)$. Furthermore, $\Vv$ is a bijection from the set of $\Jj$-unitaries onto the set of elements in the unitary group U$(2L)$ with invertible diagonal entries. In the present situation where $\Mm=\Mm^z$ for $z\in\SM^1\setminus\{-1,0,1\}$, one finds explicitly that 
\begin{equation}
\label{eq-VScat}
\Vv(\Mm^z)
\;=\;
\Ss^z
\;.
\end{equation}
Hence the scattering matrix encodes the twisted graph of the transfer matrix $\Mm^z$.
}
\hfill $\diamond$
\end{rem}

\begin{rem}
{\rm
In Definition~\ref{def-Scat}, the scattering matrix acts from the right on the Jost solutions. In the solid state physics literature of quasi-one-dimensional systems, the scattering matrix usually acts from the left on vectors. The connections between these two points of view is established by selecting two particular solutions $u^z_-\psi_+$ and $ u^{\overz}_+\phi_-$ by picking two vectors $\psi_+,\phi_-\in\CM^L$. Then the left-decreasing and right-increasing solutions $u^{\overz}_-\psi_-$ and $ u^{z}_+\phi_+$ are given by vectors $\psi_-,\phi_+\in\CM^L$ that are, according to \eqref{eq-SMatDef}, specified by
$$
\big(
u^z_- , u^{\overz}_+
\big)
\begin{pmatrix}
\psi_+ \\ \phi_-    
\end{pmatrix}
\;=\;
\big(
u^{z}_+ ,u^{\overz}_-
\big)\, 
\Ss^z
\begin{pmatrix}
\psi_+ \\ \phi_-    
\end{pmatrix}
\;=\;
\big(
u^{z}_+ , u^{\overz}_-
\big) 
\begin{pmatrix}
\phi_+ \\ \psi_-    
\end{pmatrix}
\;,
$$
namely
$$
\Ss^z
\begin{pmatrix}
\psi_+ \\ \phi_-    
\end{pmatrix}
\;=\;
\begin{pmatrix}
\phi_+ \\ \psi_-    
\end{pmatrix}
\;.
$$
Then there is a standard passage ({\it e.g.} \cite{MPK}, see Figure~\ref{fig-Schematic}) from the scattering matrix to the transfer matrix $\Mm^z$ sending the components $\psi_\pm$ on the left to those $\psi_\pm$ on the right: 
$$
\Ss^z 
\begin{pmatrix}
\psi_+ \\ \phi_-    
\end{pmatrix}
\;=\;
\begin{pmatrix}
\phi_+ \\ \psi_-    
\end{pmatrix}
\quad
\Longleftrightarrow
\quad
\Mm^z
\begin{pmatrix}
\psi_+ \\ -\psi_-    
\end{pmatrix}
\;=\;
\begin{pmatrix}
\phi_+ \\ -\phi_-    
\end{pmatrix}
\;.
$$
Hence also the plane wave transfer matrix acts from the left on the coefficients of solutions. As already stated, the link to the standard transfer matrices is explained in Section~\ref{sec-PlaneWaveRep}.
}
\hfill $\diamond$
\end{rem}

\vspace{.2cm}

Up to now, all the above follows from relatively standard techniques. The second main point of this paper is the asymptotics of the scattering matrix in the band edges $E=\pm 2$ corresponding to $z=\pm 1$. It will be shown that the limits $\lim_{z\to \pm 1} \Ss^z$ always exists so that the singularity is removable (see Section~\ref{sec-BandEdge}, in particular, Proposition~\ref{prop-MInverse}). 

\begin{theo}
\label{theo-ScatAnaly}
The scattering matrix extends analytically to $\{-1,1\}$. 
\end{theo}

Furthermore, explicit formulas for the above limits will be provided. For sake of notational simplicity, we will focus on $z=1$, as $z=-1$ can be treated similarly (or obtained by an inversion of the potential). For a generic potential, one has (see \eqref{R^1_} below)
\begin{equation}
\label{eq-Generic}
\lim_{z\to 1} \Ss^z
\;=\;
\begin{pmatrix}
0 & \one
\\
\one  & 0
\end{pmatrix}
\;.
\end{equation}
There are, however, non-generic so-called exceptional cases for which the limits of the transmission matrices are also non-trivial and can be expressed explicitly in terms of Wronskians of the Jost functions, see  Section~\ref{sec-BandEdge} below. 

\vspace{.2cm}

The second main result concerns the Levinson Theorem in the present context. Such theorems connect the number of bound and half-bound states to the winding of the scattering matrix (also called the total phase shift). For discrete Schr\"odinger operators such a connection has been found for one-dimensional operators \cite{HKS}, quasi-one-dimensional ones on a half-line \cite{AW} and higher dimensional lattice operators \cite{BSB}. It has also been understood that the equality is of topological origin \cite{KR2}, \cite{Ric}. We will set $\Ss^E=\Ss^z$ if \eqref{eq-EnergyZ} holds and $\Im m(z)>0$ and call it the on-shell scattering matrix. This reflects the fact that $\Ss^{\zb}$ is easily expressed in terms of $\Ss^z$ due to the relation $\Ss^{\zb}=\Kk(\Ss^z)^*\Kk$. The following result is proved in the Section~\ref{sec-Levinson}.	

\begin{theo}
\label{theo-Levinson}
The Hamiltonian $H$ only has a finite number $J_b$ of eigenvalues  $E_1,\ldots,E_{J_b}\in\RM$  (listed with their multiplicity) and they are outside of $[-2,2]$. Moreover, at the thresholds $E=\pm 2$, there are $J^\pm_{h}\leq L$ bounded solutions of $Hu=\pm 2 u$ which are called half-bound states. With $J_h=J_h^-+J^+_h$, one has
$$
J_b\,+\,\tfrac{1}{2}\,J_h\,-\,L
\;=\;
\int^2_{-2}\frac{d E}{2\pi\imath}\;\Tr\big((\Ss^E)^*\partial_E\Ss^E\big)
\;.
$$ 
\end{theo}

The third and last result worth mentioning in this introduction is a formula for the scattering matrix in terms of the Green function. One such connection is well-known and based on a formula for the wave operators (abelian limits) \cite{Yaf}, but here another formula is presented. It uses the tight connection between transfer matrices and Green functions for finite Jacobi matrices, also in the matrix-valued case ({\it e.g.} \cite{SB0}). This is more in the spirit of applications in mesoscopic physics \cite{Ben}, but the formula below provides the full (complex) energy dependence and hence goes beyond prior results. For $n,m\in\ZM$ and $E\not\in\sigma(H)$, the Green function is defined by
\begin{equation}
\label{eq-GreenDef}
G^E(n,m)
\;=\;
\pi_n\,(H-E)^{-1}\,(\pi_m)^*
\;\in\;\CM^{L\times L}
\;,
\end{equation}
where $\pi_n:\ell^2(\ZM,\CM^L)\to\CM^L$ is the partial isometry defined by $\pi_n \phi=\phi_n\in\CM^L$ for a vector $\phi=(\phi_n)_{n\in\ZM}\in\ell^2(\ZM,\CM^L)$. The following formula is derived in Section~\ref{sec-GreenFunction}, see equations \eqref{T^z_-} and \eqref{T^z_+}.

\begin{proposi}
\label{prop-GreenScat}
For $|z|<1$ with $z+z^{-1}\notin \sigma(H)$,  the scattering matrix $\Ss^z$ is given by
$$
\Ss^z
\;=\;
(z-z^{-1})z^{K_--K_+}
\begin{pmatrix}
G^E(K_+,K_-) &
z^{-K_+-K_-}(G^E(K_+,K_+)+\imath\nu^z)
\\
-z^{K_++K_-}(G^E(K_-,K_-)  +\imath\nu^z) &
G^E(K_-,K_+) 
\end{pmatrix}
\,.
$$
\end{proposi}

\vspace{.2cm}

In a follow-up to this paper, we will consider the scattering situation with an unbounded support of $V$, but under a short range condition. Once the existence of Jost solutions is assured by a standard argument based on the Volterra equation, many (but not all) of the results of this paper transpose to this more general situation.

\section{Transfer matrices and Jost solutions}
\label{sec-Jost}

As already stressed, the support of the potential is supposed to be finite. For sake of concreteness, let us suppose that the support is contained in $\{-K_-+1,\ldots,K_+\}$ for some finite and fixed $K_\pm$ with $K_-\leq K_+$. It is simple to construct the Jost solutions explicitly by using the transfer matrices 
$$
\Tt^E(n)
\;=\;
\begin{pmatrix}
E-V(n) & -\one \\ \one & 0
\end{pmatrix}
\;.
$$
The transfer matrices over several sites are defined to be
$$
\Tt^E(n,m)
\;=\;
\Tt^E(n)\cdots\Tt^E(m+1)
\;,
\qquad
n>m
\;,
$$
together with the convention $\Tt^E(n,n)=\one_{2L}$. Also set $\Tt^E(m,n)=\Tt^E(n,m)^{-1}$. Then one has $\Tt^E(n,n-1)=\Tt^E(n)$ and the concatenation relation $\Tt^E(n,m)\Tt^E(m,k)=\Tt^E(n,k)$. Moreover,
\begin{equation}
\label{eq-TransferComp}
\Tt^{\overline{E}}(n,m)^*\,\Ii\,\Tt^E(n,m)\;=\;\Ii\;,
\end{equation}
where $\Ii$ is second of the real Pauli matrices defined in \eqref{eq-Pauli}. For real energies $E\in\RM$, this states that the transfer matrices lie in the group $\Ii$-unitary matrices satisfying $\Tt^*\Ii\Tt=\Ii$. The transfer matrices allow to rewrite the Sch\"odinger equation $Hu=Eu$ by setting
\begin{equation}
\label{eq-PhiBuild}
\Phi(n)
\;=\;
\begin{pmatrix}
u(n+1) \\ u(n)
\end{pmatrix}
\;\in\CM^{2L\times L}
\;,
\end{equation}
namely one has
$$
\Phi(n)
\;=\;
\Tt^E(n)\,\Phi(n-1)
\;=\;
\Tt^E(n,m)\,\Phi(m)
\;.
$$
To use this iteratively, one needs some initial condition. Particular initial conditions then lead to the Jost solutions $u^z_\pm$. The suitable initial conditions are determined by the diagonalization of the unperturbed transfer matrix 
$$
\Tt_0^E
\;=\;
\begin{pmatrix} E\ & -\one \\ -\one & 0 \end{pmatrix}
\;,
$$ 
which is equal to $\Tt^E(n)$ for $n\not\in\{-K_-+1,\ldots,K_+\}$. Recalling the relation \eqref{eq-EnergyZ}, one has
$$
\begin{pmatrix}
E & -\one \\ \one & 0
\end{pmatrix}
\begin{pmatrix}
z\,\one & z^{-1}\,\one \\ \one & \one
\end{pmatrix}
\;=\;
\begin{pmatrix}
z\,\one & z^{-1}\,\one \\ \one & \one
\end{pmatrix}
\begin{pmatrix}
z\,\one & 0 \\ 0 & \overzz\,\one 
\end{pmatrix}
\;.
$$
This allows to read off the eigenvectors for the eigenvalues $z$ and $\overzz$.  Moreover, this motivates to use the notations
$$
\Cc^z
\;=\;
\begin{pmatrix}
z\,\one & \overzz\, \one \\ \one & \one
\end{pmatrix}
\;,
\qquad
\Dd^z(K)
\;=\;
\begin{pmatrix}
z^{ K}\one & 0 \\ 0 & {z}^{-K}\one 
\end{pmatrix}
\;.
$$
Now the above matrix identity can simply be written as $\Tt_0^E\Cc^z=\Cc^z\Dd^z(1)$, which shows that
$$
(\Tt_0^E)^K\,\Cc^z
\;=\;
\Cc^z\,\Dd^z(K)
\;.
$$
For the Jost solution $u^z_+$ one should therefore choose $\Phi(K_+)=\binom{z\one}{\one}$ up to a normalization factor, and for $u^z_-$ rather $\Phi(K_-)=\binom{z\one}{\one}$. Adding suitable powers of $z$ thus shows that the Jost solutions are, for $z\in\CM\setminus\{0\}$
\begin{align*}
& 
u^z_\pm(n)
\;=\;
\begin{pmatrix}
0 \\ \one
\end{pmatrix}^*
\Tt^E(n,K_\pm)
\begin{pmatrix}
z\,\one \\ \one
\end{pmatrix}
z^{K_\pm}
\,.
\end{align*}
As in \eqref{eq-PhiBuild},  the Jost solutions allow construct matrices $\Phi^z_\pm(n)\in\CM^{2L\times L}$ spanning $L$-dimensional planes in $\CM^{2L}$: 
\begin{align}
\label{eq-FrameDef0}
\Phi^z_\pm(n)
\;=\;
\begin{pmatrix}
u^z_\pm(n+1) \\ u^z_\pm(n)
\end{pmatrix}
\;=\;
\Tt^E(n,K_\pm)
\begin{pmatrix}
z\,\one \\ \one
\end{pmatrix}
z^{K_\pm}
\;.
\end{align}
Using the above notations, one has
\begin{align}
\label{eq-FundamentalDef}
\big(
\Phi^z_\pm(n),\Phi^{\overz}_\pm(n)\big)
\;=\;
\Tt^E(n,K_\pm)
\,\Cc^z\,\Dd^z(K_\pm)
\,.
\end{align}
Note that for $n\geq K_+$ one has
\begin{align}
\label{eq-FundamentalDef1}
\big(
\Phi^z_+(n),\Phi^{\overz}_+(n)\big)
\;=\;
\,\Cc^z\,\Dd^z(n)
\;=\;
\begin{pmatrix}
z^{n+1}\,\one & z^{-n-1}\, \one \\ z^n\,\one & z^{-n}\,\one
\end{pmatrix}
\;,
\end{align}
and similarly for $\big(\Phi^z_-(n),\Phi^{\overz}_-(n)\big)$ when $n\leq K_-$.

\vspace{.2cm}

To conclude this section, let us briefly point out that it is also possible to construct the linearly growing solutions $v_\pm^z$ for $z= 1$, for example:
$$
v_+^1(n)\;=\;
\begin{pmatrix}
0 \\ \one
\end{pmatrix}^*
\Tt^2(n,K_+)
\begin{pmatrix}
(K_++1)\,\one \\ K_+\,\one
\end{pmatrix}
\;.
$$

\section{Wronskian identities}
\label{sec-Wronski}

\begin{defini}
\label{def-Wronskian}
For two functions $u,v:\ZM\to\CM^{L\times L}$ and $n\in\ZM$, the Wronskian is defined by
\begin{equation}
\label{eq-WronskiDef}
W_n(u,v)
\;=\;
\imath\,\big(u(n+1)^*v(n)\,-\,u(n)^*v(n+1)\big)
\;\in\CM^{L\times L}
\;.
\end{equation}
If $W_n(u,v)$ is independent of $n$, it is simply denoted by $W(u,v)$.
\end{defini}

Clearly one has $W_n(u,v)^*=W_n(v,u)$. The definition already suggests that the Wronskian is independent of $n$ for functions $u$ and $v$ of interest. This basic fact follows from a short calculation:

\begin{lemma}
\label{lem-Wronskian}
For a pair of matrix solutions $Hu=Eu$ and $Hv=\overline{E}v$ of the Schr\"odinger equation at complex conjugate energies, the Wronskian $W_n(u,v)$ is independent of $n$. 
\end{lemma}

Thus Wronskians like $W(u_-^{\zb},u_+^{z})$ and $W(u_-^{\overzb},u_+^{z})$ do not carry the index  $n$. Moreover, they are analytic on $\mathbb{C}\setminus\{0\}$. The Wronskians of the Jost solutions can be evaluated explicitly by using the constancy in $n$ of \eqref{eq-WronskiDef}, either for $n>K_+$ or $n<K_-$. One finds the Wronskian identities
$$
W(u^{\zb}_\pm,u^z_\pm)
\;=\;0\;,
\qquad
W(u^{\overzb}_\pm,u^z_\pm)
\;=\;
(\nu^z)^{-1}\,\one
\;.
$$
Using the notations \eqref{eq-FrameDef0} allows to rewrite the Wronskian as 
\begin{equation}
\label{eq-Wronskians0}
W(u^{\zb}_+,u^z_\pm)
\;=\;
(\Phi^{\zb}_+)^*\,\tfrac{1}{\imath}\,\Ii\,\Phi^z_\pm
\;,
\qquad
W(u^{\overzb}_+,u^z_\pm)
\;=\;
(\Phi^{\overzb}_+)^*\,\tfrac{1}{\imath}\,\Ii\,\Phi^z_\pm
\;.
\end{equation}
The Wronskian identities then become
\begin{equation}
\label{eq-Wronskians}
(\Phi^{\overzb}_\pm,\Phi^{\zb}_\pm)^*\,\tfrac{1}{\imath}\,\Ii\,(\Phi^z_\pm,\Phi^{\overz}_\pm)\;=\;(\nu^z)^{-1}\,\Jj
\;.
\end{equation}
Here the index $n$ in all $\Phi^z_\pm(n)$ is dropped. In particular, these identities imply that the matrices $(\Phi^z_\pm,\Phi^{\overz}_\pm)\in\CM^{2L\times 2L}$ are invertible. 

\vspace{.2cm}

The following Wronskian identity involving the derivatives of the Jost solutions w.r.t. $z$ will also be used below.

\begin{lemma}
\label{lem-Wronskian2}
For $z\in\CM\setminus\{0\}$, $\sigma,\eta\in\{-,+\}$ and $n\in\ZM$, one has
$$
W_n(u^{\zb}_\sigma,\partial_z u^z_\eta)
\;=\;
W_{n-1}(u^{\zb}_\sigma,\partial_z u^z_\eta)
\;-\;
\imath\, (1-z^{-2})\,u^{\zb}_\sigma(n)^* u^z_\eta(n)
\;.
$$
A similar identity holds for $W_n(u^{\overzb}_\sigma,\partial_z u^z_\eta)$.
\end{lemma}

\noindent {\bf Proof.} The equation $Hu^z_\pm=(z+z^{-1})u^z_\pm$ at every point $n\in\ZM$ reads
\begin{equation}
\label{eq-WronkPrep1}
u^z_\sigma(n+1)\,+\,V(n) u^z_\sigma(n)\,+\,u^z_\sigma(n-1)
\;=\;
(z+z^{-1})u^z_\sigma(n)
\;.
\end{equation}
Taking the adjoint of this equation with $z$ replaced by $\zb$ and multiplying from the right by $\partial_z u^z_\eta(n)$ leads to
\begin{equation}
\label{eq-WronkPrep2}
u^{\zb}_\sigma(n+1)^*\partial_z u^z_\eta(n)\,+\, u^{\zb}_\sigma(n)^*V(n)\partial_z u^z_\eta(n)\,+\,u^{\zb}_\sigma(n-1)^*\partial_z u^z_\eta(n)
\;=\;
(z+z^{-1})u^{\zb}_\sigma(n)^*\partial_z u^z_\eta(n)
\;.
\end{equation}
Deriving equation \eqref{eq-WronkPrep1} for $\sigma$ replaced by $\eta$ w.r.t. $z$ gives
\begin{equation}
\label{eq-WronkPrep3}
\partial_z u^z_\eta(n+1)\,+\,V(n) \partial_z u^z_\eta(n)\,+\,\partial_z u^z_\eta(n-1)
\;=\;
(1-z^{-2})u^z_\eta(n)
\,+\,(z+z^{-1})\partial_z u^z_\eta(n)
\;.
\end{equation}
Multiplying \eqref{eq-WronkPrep3} from the left by $u^{\zb}_\sigma(n)^*$ and then subtracting \eqref{eq-WronkPrep2} leads to the claim.
\hfill $\Box$

\section{Plane wave transfer matrices}
\label{sec-PlaneWaveRep}

The plane wave transfer matrix $\Mm^z$ is introduced in Definition~\ref{def-Mtransfer}. This terminology will be explained and justified in this section. Recall the definition \eqref{eq-MMatDef},  namely $(u^z_- , u^{\overz}_-)=(u^{z}_+ , u^{\overz}_+)\Mm^z$. Using the matrices $\Phi^z_\pm(n)\in\CM^{2L\times L}$ in \eqref{eq-FrameDef0}, this can be rewritten as
$$
\big(
\Phi^z_-,\Phi^{\overz}_-\big)
\;=\;
\big(
\Phi^z_+,\Phi^{\overz}_+\big)
\,
\Mm^z
\;.
$$
Note that on both sides of this equality one can still replace the lattice site $n$. Multiplying the equation from the left with $(\Phi^{\overzb}_\pm,\Phi^{\zb}_\pm)^*\,\tfrac{1}{\imath}\,\Ii$ and taking into account the Wronskian identities \eqref{eq-Wronskians} as well as $\Jj^2=\one$ shows
\begin{equation}
\label{eq-MDef}
\Mm^z
\;=\;
\nu^z\;\Jj\,\big(\Phi^{\overzb}_+,\Phi^{\zb}_+\big)^*\,\tfrac{1}{\imath}\,\Ii\,\big(\Phi^z_-,\Phi^{\overz}_-\big)
\;.
\end{equation}
It follows that $\Mm^z$ is analytic in $z$ away from $\{-1,0,1\}$. In order to make connections with the transfer matrix, let us now use \eqref{eq-FundamentalDef} for the minus sign and $n=K_+$ followed by \eqref{eq-FundamentalDef1}:
\begin{align*}
\Tt^E(K_+,K_-)
\,\Cc^z\,\Dd^z(K_-)
& 
\;=\;
\big(
\Phi^z_-(K_+),\Phi^{\overz}_-(K_+)\big)
\\
&
\;=\;
\big(
\Phi^z_+(K_+),\Phi^{\overz}_+(K_+)\big)
\,\Mm^z
\\
&
\;=\;
\Cc^z\,\Dd^z(K_+)\,\Mm^z
\;,
\end{align*}
implying
\begin{equation}
\label{eq-TransferMat}
\Mm^z
\;=\;
\big(\Cc^z\Dd^z(K_+)\big)^{-1}\,
\Tt^E(K_+,K_-)
\,\big(\Cc^z\Dd^z(K_-)\big)
\;.
\end{equation}
This suggests introducing the one-step plane wave transfer matrices at $n\in\ZM$ by
\begin{equation}
\label{eq-TransferMatSingle}
\Mm^z(n)
\;=\;
\big(\Cc^z\Dd^z(n)\big)^{-1}\,
\Tt^E(n)
\,\big(\Cc^z\Dd^z(n-1)\big)
\;,
\end{equation}
as well as the several step version by $\Mm^z(n,m)=\Mm^z(n)\cdots\Mm^z(m+1)$  and $\Mm^z(m,n)=\Mm^z(n,m)^{-1}$ for $n>m$, just as for the transfer matrices. Also let us set $\Mm^z(n,n)=\one$. With these notations, one has $\Mm^z(n,n-1)=\Mm^z(n)$ and the transfer matrix of  Definition~\ref{def-Mtransfer} is $\Mm^z=\Mm^z(K_+,K_-)$. One can now deduce a first crucial feature of $\Mm^z(n,m)$, namely their multiplicativity still holds. Indeed, decomposing $\Tt^E( n, m)=\Tt^E( n)\Tt^E( n-1, m)$ leads to:
\begin{align}
&
{\Mm}^z( n, m)
\nonumber
\\
&
\;=\;
\big(\Cc^z\Dd^z(n)\big)^{-1}\,
\Tt^E( n)\,
\big(\Cc^z\Dd^z(n-1)\big)\,
\big(\Cc^z\Dd^z(n-1)\big)^{-1}
\,\Tt^E(n-1, m)
\,(\Cc^z\Dd^z(m)\big)
\nonumber
\\
&
\;=\;
{\Mm}^z( n)\,
{\Mm}^z(n-1, m)
\nonumber
\\
&
\;=\;
{\Mm}^z( n)\cdots
{\Mm}^z( m+1)
\;.
\label{eq-TransferMProp}
\end{align}
From \eqref{eq-TransferMatSingle}, one can also calculate the plane wave transfer matrix explicitly:
\begin{equation}
\label{eq-TransferMExplicit}
{\Mm}^z(n)
\;=\;
\begin{pmatrix}
\one & 0 \\ 0 & \one 
\end{pmatrix}
\;+\;
\imath\,\nu^z\,
\begin{pmatrix}
V(n) & z^{-2n}V(n) \\ -z^{2n} V(n) & -V(n) 
\end{pmatrix}\,
\;.
\end{equation}
Therefore ${\Mm}^z(n)=\one$ if the potential $V(n)$ vanished. From \eqref{eq-TransferMExplicit} one readily checks the identities ${\Mm}^{\overzb}(n)^*\Jj{\Mm}^z(n)=\Jj$ and ${\Mm}^{z}(n)\Jj{\Mm}^{\overzb}(n)^*=\Jj$. Combined with \eqref{eq-TransferMProp} one deduces the second set of crucial properties of the plane wave transfer matrices:
\begin{align}
\label{eq-MJunitarity}
& \Mm^{\overzb}( n,m)^*\,\Jj\,\Mm^z(n,m)
\;=\;
\Jj\;,
\\
&
\Mm^{z}(n,m)\,\Jj\,\Mm^{\overzb}(n,m)^*
\;=\;
\Jj\;.
\label{eq-MJunitarity2}
\end{align}
For $z\in\SM^1\setminus\{-1,1\}$, this shows that  $\Mm^z(n,m)$ is $\Jj$-unitarity, namely it satisfies $\Mm^*\Jj\Mm=\Jj$.  This also implies \eqref{eq-MJunitarityIntro}. 

\vspace{.2cm}

In conclusion, the matrices $\Mm^{z}(n,m)$ have the multiplicativity property and are $\Jj$-unitarity, similar as the transfer matrices which are multiplicative and $\Ii$-unitary. Furthermore, they are trivial for vanishing potential and are thus adapted to plane waves. All this justifies the terminology used. The link to the transfer matrices $\Tt^E(n,m)$ is established by \eqref{eq-TransferMatSingle}. This shows that the passage from $\Tt^E(n,m)$ to $\Mm^{z}(n,m)$ is given by the basis change induced by $\Cc^z$, followed by the site-dependent diagonal factors (which destroy the representation property). The basis change $\Cc^z$ for $z=\pm \imath$ induces via the M\"obius action the Cayley transformation and this is well-known to induce a map from $\Ii$-unitaries to $\Jj$-unitaries ({\it e.g.} \cite{SB}). This is also effect of the basis change here. For sake of concreteness,  let us write out $\Mm^z( n, m)$ in terms of the $L\times L$ block entries (depending on $E$) of the transfer matrix:
\begin{equation} 
\label{eq-TransferCoeff}
\Tt^E(n, m)
\;=\;
\begin{pmatrix}
A & B \\ C & D
\end{pmatrix}
\;.
\end{equation}
One finds 
$$
\Mm^z(n, m)
\,=\,
\imath\,\nu^z
\Dd^z( n)^{-1}
\begin{pmatrix}
-B+C-zA+\overzz D  & -B+z^{-1}(D-A)+z^{-2} C 
\\  
B+z(A-D)-{z}^{2} C & -C+B+z^{-1} A-{z}D  
 \end{pmatrix}
\Dd^z(m)
\;.
$$

Next let us look at real $z\in\RM$, corresponding to energies $E\in\RM\setminus(-2,2)$.  It is a matter of (tedious) calculation to check that
\begin{equation}
\label{eq-MIunitary}
\Mm^z(n)^*\,\Ii\,\Mm^z(n)
\;=\;\Ii
\;,
\end{equation}
which is the $\Ii$-unitary of $\Mm^z(n)$ for real $z\in\RM$. This implies that also $\Mm^z(n,m)$ and thus $\Mm^z$ is $\Ii$-unitary, which is item (ii) of Proposition~\ref{prop-TransScatProp}.

\section{Matrix entries of plane wave transfer matrix}
\label{sec-MatrixEntries}

Let us begin by recalling the defining relation \eqref{eq-MDef3} of $\Mm^z$, namely  $\big( u^z_- ,u^{\overz}_-\big)=\big(u^{z}_+ , u^{\overz}_+ \big) \Mm^z$. Multiplying  this by the inverse of $\Mm^z$ leads to $\big(u^{z}_+ , u^{\overz}_+ \big)=\big( u^z_- ,u^{\overz}_-\big)(\Mm^z)^{-1}$. Introducing notations for the matrix coefficients
%
$$
(\Mm^z)^{-1}
\;=\;
\begin{pmatrix}
M^z_+ & N^{{\overz}}_+ \\
N^{z}_+ & M^{\overz}_+
\end{pmatrix}
\;,
$$
the above two equations together with \eqref{eq-MDef3} imply
\begin{equation}
\label{eq-udecomp}
u_+^z
\;=\;u_-^zM^z_+\;+\;u_-^{\overz}N^z_+
\;,
\qquad
u_-^{\overz}
\;=\;u_+^zN^z_-\;+\;u_+^{\overz}M^z_-
\;,
\end{equation}
holding for $z\in\CM\setminus\{-1,0,1\}$. This can also be rewritten as
$$
\Phi_+^z
\;=\;\Phi_-^zM^z_+\;+\;\Phi_-^{\overz}N^z_+
\;,
\qquad
\Phi_-^{\overz}
\;=\;\Phi_+^zN^z_-\;+\;\Phi_+^{\overz}M^z_-
\;.
$$ 
Therefore the Wronskian identities \eqref{eq-Wronskians} lead to
\begin{align}
\label{eq-MNId0}
\begin{split}
& 
M^z_+
\;=\;
\nu^z\,
(\Phi_-^{\overzb})^*\,\tfrac{1}{\imath}\,\Ii\,\Phi_+^z
\;=\;
\nu^z\,
W(u_-^{\overzb},u^z_+)
\;,
\\
&
N^z_+
\;=\;
-\,
\nu^z\,
(\Phi_-^{\zb})^*\,\tfrac{1}{\imath}\,\Ii\,\Phi_+^z
\;=\;
-\,\nu^z\,
W(u_-^{\zb},u^z_+)
\;,
\\
&
N^z_-
\;=\;
\nu^z\,
(\Phi_+^{\overzb})^*\,\tfrac{1}{\imath}\,\Ii\,\Phi_-^{\overz}
\;=\;
\nu^z\,
W(u_+^{\overzb},u^{\overz}_-)
\;,
\\
& M^z_-
\;=\;
-\,\nu^z\,(\Phi_+^{\zb})^*\,\tfrac{1}{\imath}\,\Ii\,\Phi_-^{\overz}
\;=\;
-\,\nu^z\,
W(u_+^{\zb},u^{\overz}_-)
\;.
\end{split}
\end{align}
This shows that $M^z_\pm$ and $N^z_\pm$ are analytic on $\CM\setminus\{-1,0,1\}$.  
Hence $\Mm^z$ can also be written using the Wronskians:
\begin{equation}
\label{eq-MDef2}
\Mm^z
\;=\;
\nu^z\;
\begin{pmatrix}
W(u^{\overzb}_+,u^z_-)   & W(u^{\overzb}_+,u^{\overz}_-)  
\\
-W(u^{\zb}_+,u^z_- )  &  -\,W(u^{\zb}_+,u^{\overz}_-)  
\end{pmatrix}
\;.
\end{equation}
Next let us note that $(\Mm^z)^{-1}=\Jj(\Mm^{\overzb})^*\Jj$ due to \eqref{eq-MJunitarity2}, which is equivalent to
\begin{equation}
(N^z_+)^*
\;=\;
-\,N^{\overzb}_-\;,
\qquad
(M^z_+)^*
\;=\;
M^{\zb}_-
\;.
\label{eq-MNId2}
\end{equation}
%
%
Furthermore, writing out the relations \eqref{eq-MJunitarity} and \eqref{eq-MJunitarity2}, a computation leads to:
\begin{align}
\label{eq-MNId3a}
&
M^{z}_+(M^{\overzb}_+)^*\;=\;\one\,+\,N^{\overz}_+(N^{\zb}_+)^*\;,
\qquad
(M^{\overzb}_-)^*M^z_-\;=\;\one\,+\,(N^{\overzb}_-)^*N^z_-\;,
\\
&
M^{z}_+N^z_-\;=\;-\,N^{\overz}_+M^z_-\;,
\qquad\qquad\qquad\,
M^{z}_-N^{z}_+\;=\;-\,N^{\overz}_-M^z_+\;,
\label{eq-MNId3b}
\\
&
(M^{\overzb}_+)^*M^{z}_+\;=\;\one\,+\,(N^{\overzb}_+)^*N^z_+\;,
\qquad
M^z_-(M^{\overzb}_-)^*\;=\;\one\,+\,N^{\overz}_-(N^{\zb}_-)^*\;.
\label{eq-MNId3c}
\end{align}
Combined with \eqref{eq-MNId2} these equations imply for $z\in\RM$:
\begin{equation}
\label{eq-MNId4}
(N^z_-)^*M^z_-\;=\;(M^z_-)^*N^z_-
\;,
\qquad
(M^{\overz}_-)^*M^z_-\;=\;\one \,+\,(N^{\overz}_-)^*N^{z}_-
\;.
\end{equation}
%

\begin{lemma}
\label{lem-Minvertible}
For $z\in\SM^1\setminus\{-1,1\}$, $M^z_\pm$ are invertible. 
\end{lemma}

\noindent {\bf Proof.}
For $z\in\SM^1$, one has $\overzb=z$ and hence the identities \eqref{eq-MNId3c} imply $(M^z_\pm)^*M^z_\pm\geq \one$ so that $M^z_\pm$ are invertible. 
\hfill $\Box$

\section{Scattering matrix}
\label{sec-Scat}

The scattering matrix was introduced in Definition~\ref{def-Scat}. This section merely expresses the scattering matrix in terms of the matrix coefficients $M^z_\pm$ and $N^z_\pm$, and then deduces some first basic properties. For that purpose, let $z\in\CM_0$ so that the inverses $(M^z_\pm)^{-1}$ exist. Then one can rewrite \eqref{eq-udecomp} as
$$
u_-^z\;=\;u_+^z(M^z_+)^{-1}\;-\; u_-^{\overz}N^z_+  (M^z_+)^{-1}
\;,
\qquad
u_+^{\overz}\;=\;u_-^{\overz}(M^z_-)^{-1}\;-\;u_+^zN^z_-(M^z_-)^{-1}
\;.
$$
Comparing with \eqref{eq-SMatDef} in Definition~\ref{def-Scat}, the scattering matrix at $z\in\CM_0$ thus is
\begin{equation}
\label{eq-ScatMN}
\Ss^z
\;=\;
\begin{pmatrix}
(M^z_+)^{-1} & - N^z_-(M^z_-)^{-1} \\
-N^z_+(M^z_+)^{-1} & (M^z_-)^{-1}
\end{pmatrix}
\;,
\end{equation}
which allows to read off the transmission and reflection coefficients.  Furthermore the equations \eqref{eq-MNId2} and \eqref{eq-MNId3b} also allow to rewrite  \eqref{eq-ScatMN}, {\it e.g.}
\begin{equation}
\label{eq-ScatMN2}
\Ss^z
\;=\;
\begin{pmatrix}
((M^{\zb}_-)^*)^{-1}  & - N^z_-(M^z_-)^{-1} \\
(M^z_-)^{-1}N^{\overz}_- & (M^z_-)^{-1}
\end{pmatrix}
\;.
\end{equation}
Based on these formulas  \eqref{eq-MNId2} to \eqref{eq-MNId3c}, a calculation allows to verify the claims of Proposition~\ref{prop-TransScatProp}(ii). Furthermore, Proposition~\ref{prop-TransScatProp}(v) also follows because $z\mapsto M^z_-$ is meromorphic due to \eqref{eq-MNId0}) and therefore $z\mapsto\det(M^z_-)$ also. Moreover, $M_{\pm}^z \to \one$ for $ z \to 0$, and the limits $(M_{\pm}^z)^{-1}$ exist when $z \to \pm1$ (see Propositions~\ref{prop-HighEnergy} and \ref{prop-MInverse}).  Hence the zeros of the latter map and thus points of singularity of $(M^z_-)^{-1}$ form a discrete set. Hence also $z\mapsto (M^z_-)^{-1}$ is meromorphic and so is $z\mapsto \Ss^z$.

\section{Bound states}
\label{sec-BoundStates}

In this section, $E\in\RM\setminus[-2,2]$. Hence $z\in \RM\setminus\{-1,0,1\}$.  If $|z|<1$, then the Jost solution $u^z_+$ is decaying exponentially at $+\infty$, and the Jost solution $u^{\overz}_-$ decays exponentially at $-\infty$. If these two solutions match together, one gets eigenstates for the selfadjoint operator $H$. Of course, such eigenstates cannot appear for complex energies. 

\begin{proposi}
\label{prop-MComplexInvert}
For $E=z+\overzz\not\in\sigma(H)$ with $|z|<1$, $M^z_\pm$ are invertible.
\end{proposi}

\noindent {\bf Proof.} Recall from \eqref{eq-MMatDef} that $u_-^{\overz}=u_+^zN^z_-+u_+^{\overz}M^z_-$. Let us assume that $M_{\pm}^z$ is not invertible and then show that $E\in \sigma(H)$. If $M^z_-$ has a non-trivial kernel, then the left and right solutions match on a subspace which produces square integrable eigenstates, so that indeed $E\in \sigma(H)$. Now the invertibility of $M_+^z$ follows from  \eqref{eq-MNId2}.
\hfill $\Box$

\vspace{.2cm}

Let us now look more closely at the multiplicity of the eigenvalues. The intersection of the space of left decreasing and right decreasing solution lead to square integrable bound state at energy $E$. More precisely, this intersection can be parametrized by
\begin{equation}
\label{eq-IntersecGen}
\Ran\big(\Phi^z_+(K_+)\big)\cap \Ran\big(\Tt^E(K_+,K_-)\Phi^{\overz}_-(K_-)\big)
\;.
\end{equation}
Provided the intersection is non-trivial, one can then choose $\phi\in\CM^L$  such that $\Phi^z_+(K_+)\phi$ is in the intersection, and this provides the bound state
$$
u_\phi(n)\;=\;\binom{0}{\one}^*\Tt^E(n,K_+)\Phi^z_+(K_+)\phi
\;.
$$
%

\begin{rem}
\label{N_iso}
{\rm
Replacing \eqref{eq-MNId2} into \eqref{eq-MNId3a}, one finds
$$
(M_-^{\overzb})^*M_-^z
\;=\;\one\,-\,N_+^zN_-^z
\;,, 
\qquad 
(M^{\overzb}_+)^*M^{z}_+\;=\;\one\,-\,N_-^zN_+^z
\;.
$$ 
This implies that $N_+^zN_-^z\phi = \phi$ for $\phi \in \Ker(M_-^z)$, and  $N_-^zN_+^z\phi = \phi$ for $\phi \in \Ker(M_+^z)$. Moreover, equations \eqref{eq-MNId3b} imply that $N_-^z(\Ker(M_-^z))\subset \Ker(M_+^z)$ and $N_+^z(\Ker(M_+^z))\subset \Ker(M_-^z)$. Therefore 
\begin{equation}
\label{eq-Isomorphism}
N_-^z\big|_{\mbox{\rm\small Ker}(M_-^z)}\;:\;\Ker(M_-^z)\to \Ker(M_+^z)
\end{equation}
is an isomorphism with inverse $N_+^z|_{\mbox{\rm\small Ker}(M_+^z)}$.
}
\hfill $\diamond$
\end{rem}

\begin{proposi}
\label{prop-BoundStates}
For $E=z+z^{-1}\in\RM$ with $|z|<1$, one has
\begin{align*}
\mbox{\rm multiplicity of }E\;\mbox{\rm as eigenvalue of }H
&
\;=\;
\dim\,\Ker(M^z_\pm)
\\
&
\;=\;
\mbox{\rm order of $z$ as zero of }
z'\mapsto\det(M_\pm^{z'})
\;.
\end{align*}
In particular, $M^z_\pm$ is invertible if $z+\overzz$ is not an eigenvalue.
\end{proposi}

\noindent {\bf Proof.} The argument in the proof of Proposition~\ref{prop-MComplexInvert} and the discussion above imply the first equality. For the proof of the second equality let us notice that $z+1/z \in \mathbb{R}$ with $|z|<1$. This implies that $z\in \mathbb{R}$. Let $p$ denote  the dimension of $\Ker(M_-^{z})$ and let $\{w_1,\ldots,w_p\}$ be a basis of $\Ker(M_-^{z})$. Since $N_-^z:\Ker(M_-^z) \to \Ker(M_+^z)$ is an isomorphism by Remark \ref{N_iso}, it follows that $\{N_-^zw_1,\ldots,N_-^zw_p\}$ is a basis of $\Ker(M_+^z) = \Ker((M_-^z)^*)=\Ran(M_-^z)^\perp$ (see \eqref{eq-MNId2}). Let $\{v_{p+1},\ldots,v_{L}\}$ be an orthonormal basis of $\Ran(M_-^z)$ and $\{u_{p+1},\ldots,u_{L}\}$ vectors such that $M_-^ru_j=v_j$ for $j \in \{p+1,\ldots,L\}.$ Then $\{w_1,\ldots,w_p,u_{p+1},\ldots,u_{L}\}$ and $\{N_-^zw_1,\ldots,N_-^zw_p,v_{p+1},\ldots,v_{L}\}$ are bases of $\mathbb{C}^L$. Let us introduce the invertible $L\times L$ matrices
$$
U_1\;=\;
\big(
w_1,\ldots , w_p , u_{p+1} ,\ldots , u_{L}
\big)
\;,
\qquad 
V_1
\;=\;
\big(
N_-^zw_1 ,  \ldots , N_-^zw_p ,v_{p+1} ,\ldots , v_{L}
\big)
\;.
$$
It follows that 
$$
V_1^*M_-^zU_1
\;=\;
\begin{pmatrix}
0 & 0 \\
0 & \mathbf{1}
\end{pmatrix}
\;.
$$
Since $\zeta \mapsto M_-^{\zeta}$ is analytic around $z$, one has $M_-^{\zeta}=M_-^z+(\zeta-z)\partial_z M_-^z+\Oo((\zeta-z)^2)$ so that
$$
V^*M_-^{\zeta}U
\;=\;
\begin{pmatrix}
(\zeta-z)V_2^*\partial_z M_-^zU_2  & (\zeta-z)B \\
(\zeta-z)C & \mathbf{1}+(\zeta-z)D 
\end{pmatrix}
\;+\;
\Oo\big((\zeta-z)^2\big)
\;,
$$ 
for some constant matrices $B,C,D$ and where $U_2, V_2$ are given by 
$$
U_2\;=\;
\big(
w_1 , \ldots , w_p
\big)
\;,
\qquad
V_2\;=\;
\big(
N_-^zw_1 , \ldots , N_-^zw_p 
\big)
\;=\;
N_-^zU_2
\;.
$$
The matrix $V_2^*\partial_zM_-^zU_2$ is invertible because Lemma \ref{lem-positive} implies that for $z>0$ and all $\phi \in \mathbb{C}^p$
$$
\phi^*V_2^*\partial_zM_-^zU_2\phi
\;=\;
(V_2\phi)^*\partial_zM_-^zU_2\phi 
\;=\; 
(U_2\phi)^*(N_-^z)^*\partial_z{M_-^z}(U_2\phi)
\;>\;
0
\;,
$$
and similarly for $z<0$. Using the Schur complement formula for the determinant, one obtains
\begin{align*}
\det(V_1^*M_-^{\zeta}U_1)
&
\;=\;
\det\!\big(\mathbf{1}+(\zeta-z)D+\Oo((\zeta-z)^2)\big)\,\det\!\big((\zeta-z)(V_2^*\partial_z M_-^zU_2 + \Oo(\zeta-z))\big)
\\
&
\;=\;
(\zeta-z)^pg(\zeta)
\;,
\end{align*}
with $g$ being a function satisfying $g(z)=\det(V_2^*\partial_z M_-^zU_2) \not = 0$. This implies the claim.
\hfill $\Box$

\begin{lemma}
\label{lem-positive} For $z\in (-1,1)$ and $\phi\in\Ker(M^z_-)$,
$$
\phi^*\,(N_-^z)^*\partial_z M^z_- \phi
\;=\;
z^{-1}\,\|u^{\overz}_-\phi\|^2
\;.
$$
\end{lemma}

\noindent {\bf Proof:} First of all note that for $\phi\in\Ker(M^z_-)$, \eqref{eq-udecomp} implies that $u_-^{\overz}\phi =u_+^zN^z_-\phi$ which is hence a square summable vector both at $-\infty$ and $\infty$. Consequently the $\ell^2$-norm $\|u^{\overz}_-\phi\|$ appearing in the statement is indeed finite.

\vspace{.1cm}

Let us start from $M^z_-=-\nu^z W(u_+^{z},u^{\overz}_-)$ given in
\eqref{eq-MNId0} with $\zb=z\in\RM$. Deriving leads to
$$
\partial_z M^z_-
\;=\;
-\,(\partial_z\nu^z) W(u_+^{z},u^{\overz}_-)
\,-\,\nu^z W_n(\partial_{z}u_+^{z},u^{\overz}_-)
\,-\,\nu^z W_n(u_+^{z},\partial_zu^{\overz}_-)
\;,
$$
where $n\in\ZM$ is arbitrary. As $W(u_+^{z},u^{\overz}_-)\phi=0$ and $u_-^{\overz}\phi =u_+^zN^z_-\phi$, this implies
\begin{align*}
\phi^*(N^z_-)^*\partial_z M^z_-\phi
&
\;=\;
-\,\nu^z \phi^*(N^z_-)^* W_n(\partial_{z}u_+^{z},u^{\overz}_-)\phi
\,-\, \nu^z \phi^*(N^z_-)^*  W_n(u_+^{z},\partial_zu^{\overz}_-)\phi
\\
&
\;=\;
-\,\nu^z \phi^*(N^z_-)^* W_n(\partial_{z}u_+^{z},u^{z}_+)N^z_-\phi
\,-\, \nu^z \phi^* \,W_n(u_-^{\overz},\partial_zu^{\overz}_-)\phi
\;.
\end{align*}
As $|z|<1$, one has
$$
\lim_{n\to\infty}\,W_n(\partial_{z}u_+^{z},u^{z}_+)\;=\;0
\;.
$$
To compute the limit of the other contribution let us invoke Lemma~\ref{lem-Wronskian2} iteratively. As $\partial_zf(1/z)=-z^{-2}\partial_{\overz}f(1/z)$ for every analytic function, one computes
$$
\phi^*(N^z_-)^*\partial_z M^z_-\phi
\,=\,
z^{-2}\nu^z  
\lim_{n\to\infty}
\left(
\phi^* W_k(u_-^{\overz},\partial_{\overz}u^{\overz}_-)\phi
-
\!\!
\sum_{m=k+1}^n
\!\!
\imath(1-z^{2})\,\phi^*u^{\overz}_-(m)^* u^{\overz}_-(m)\phi
\right)
.
$$
This holds for any $k$. As
$$
\lim_{k\to-\infty}\;
W_k(u_-^{\overz},\partial_{\overz}u^{\overz}_-)
\;=\;0
\;,
$$
the claim now follows in the limit $k\to-\infty$.
\hfill $\Box$

\begin{coro}
\label{coro-C0}
Recall that $\CM_0$ is the set of points $z\in\CM\setminus\{-1,0,1\}$ where $M^z_\pm$ are invertible. The set $\CM_0$ contains an open neighborhood of the unit disc with the points $\{-1,0,1\}$ and $\{z\in\CM\;:\; z+z^{-1}\in\sigma_p(H)\}$ removed. 
\end{coro}

\noindent {\bf Proof.} For $z\in\SM^1\setminus\{-1,1\}$ the invertibility of $M^z_-$  is stated in Lemma~\ref{lem-Minvertible}. For $\Im m(z)\not=0$ this is Proposition~\ref{prop-MComplexInvert}. For real $z\not = 0$ the invertibility is characterized in Proposition~\ref{prop-BoundStates}. 
\hfill $\Box$

\vspace{.2cm}

Corollary~\ref{coro-C0} contains no statement about the analyticity of $M^z_-$ in $\{-1,0,1\}$. Section~\ref{sec-HighEnergy} will consider $z=0$ and Sections~\ref{sec-BandEdge} and \ref{sec-AsympConst} the points $z=\pm 1$. The remainder of this section consists of comments about geometric structures behind Proposition~\ref{prop-BoundStates}. They freely use the notion of intersection theory of Lagrangian planes and some of the terminology linked to the theory of the Maslov index, see {\it e.g.} \cite{SB,SB2}. Let us stress that these remarks are not relevant for the following though.

\begin{rem}
\label{sec-GeoArg}
{\rm
This remark provides alternative proofs of the two identities in Proposition~\ref{prop-BoundStates}. 
The intersection \eqref{eq-IntersecGen} is actually the intersection of two $\Ii$-Lagrangian planes and hence its dimension $J_z$ can be calculated by intersection theory ({\it e.g.} Proposition~2 in \cite{SB2}) as  
\begin{align*}
J_z
&
\;=\;
\dim\;\Ker
\left(
\Phi^z_+(K_+)^*\, \Ii\,\Tt^E(K_+,K_-)\Phi^{\overz}_-(K_-)
\right)
\\
&
\;=\;
\dim\;\Ker
\left(
\nu^z\;\Phi^z_+(K_+)^*\, \tfrac{1}{\imath}\,\Ii\,\Phi^{\overz}_-(K_+)
\right)
\\
&
\;=\;
\dim\;\Ker
\left(
\binom{0}{\one}^*\Mm^z\binom{0}{\one}
\right)
\;=\;
\dim\;\Ker
\left(
M^z_-
\right)
\;,
\end{align*}
where \eqref{eq-MDef} was used. This shows the first identity. For the second equality,   let us recall from \eqref{eq-MNId2} that $(M^z_+)^*=M^{\zb}_-$. Hence the analytic matrix function
$$
\Hh^{z}
\;=\;
\begin{pmatrix}
0 & M^{z}_- \\ M^{z}_+ & 0
\end{pmatrix}
$$
is selfadjoint for $z\in(-1,1)$. Therefore by Rellich's analytic perturbation theory the eigenvalue functions can be labelled (at level crossings) to that they are all real analytic on $(-1,1)$. Thus also $z\in(-1,1)\mapsto \det(\Hh^{z})=\det(M^{z}_-)\det( M^{z}_+)=|\det(M^{z}_-)|^2$  has a zero of order at least $2\,\dim(\Ker(M^z_-))=2\,\dim(\Ker(M^z_+))$. To verify that the zero is of this order $2\,\dim(\Ker(M^z_\pm))$, it will be shown that the restriction of the selfadjoint matrix $\partial_z \Hh^{z}$ to $\Ker(\Hh^z)=\Ker(M^z_+)\oplus\Ker(M^z_-)$ is non-degenerate. This follows from the Max-Min principle by showing that $\Ker(\Hh^z)$ has two subspaces of dimension $\dim(\Ker(M^z_\pm))$, on one of which $\partial_z \Hh^{z}$ is positive definite and on the other of which it is negative definite.

\vspace{.1cm}

For this purpose, let us recall from the remark above that $N^z_-:\Ker(M^z_-)\to\Ker((M^z_-)^*)=\Ker(M^z_+)$ is an isomorphism. Hence for $\phi\in\Ker(M^z_-)$, one has $N^z_-\phi\in \Ker(M^z_+)$ and therefore $N^z_-\phi\oplus \sigma \phi \in\Ker(\Hh^z)$ for $\sigma\in\{-1,1\}$. Now
\begin{align*}
\begin{pmatrix}
N^z_-\phi \\ \sigma \phi 
\end{pmatrix}^*
\partial_z\Hh^z
\begin{pmatrix}
N^z_-\phi \\ \sigma \phi 
\end{pmatrix}
&
\;=\;
\sigma\,\phi^*\,(N_-^z)^*\partial_z M^z_- \phi
\;+\;
\sigma\,\phi^*\,\partial_z M^z_+ N_-^z\phi
\\
&
\;=\;
2\,\sigma\,\phi^*\,(N_-^z)^*\partial_z M^z_- \phi
\;,
\end{align*}
where the second equality follows from \eqref{eq-MNId2} and the selfadjointness of $(N^z_-)^*M^z_-$, see \eqref{eq-MNId4}. Therefore the proof of the the second equality is again completed by Lemma~\ref{lem-positive}.
}
\hfill $\diamond$
\end{rem}

\begin{rem}
{\rm  For $z\in(-1,0)\cup(0,1)$, Proposition~\ref{prop-TransScatProp}(ii) or equivalently \eqref{eq-MNId4} imply that the column vectors of $\binom{N^z_-}{M^z_-}$ span an $\Ii$-Lagrangian plane in $\CM^{2L}$, namely an $L$-dimensional subspace on which $\Ii$ viewed as sesquilinear quadratic form vanishes. One can then introduce the phase of this plane via the stereographic projection:
$$
U^z\;=\;(N^z_-\,-\,\imath\,M^z_-)(N^z_-\,+\,\imath\,M^z_-)^{-1}
\;.
$$
The inverse of $N^z_-+\imath\,M^z_-$ indeed exists and $U^z$ is unitary. Now according to Proposition~\ref{prop-BoundStates}, the multiplicity of $E=z+z^{-1}$ as eigenvalue of $H$ is given by the dimension of the intersection of the two $\Ii$-Lagrangian planes $\binom{N^z_-}{M^z_-}$ and $\binom{\one}{0}$. This dimension is equal the dimension of $z=1$ as eigenvalue of $U^z$ because
$$
\psi=(N^z_-\,+\,\imath\,M^z_-)^{-1}\phi\,\in\,\Ker(U^z-\one)
\quad
\Longleftrightarrow
\quad
\phi\,\in\,\Ker(M^z_-)
\;.
$$
Furthermore, the identity
$$
\imath\,(U^z)^*\partial_z U^z
\;=\;
-2\,\big((N^z_-\,+\,\imath\,M^z_-)^{-1}\big)^*
\big((N^z_-)^* \partial_z M^z_-
-(M^z_-)^*\partial_z N^z_-\big)
(N^z_-\,+\,\imath\,M^z_-)^{-1}
$$
combined with Lemma~\ref{lem-positive} shows that for all $\psi=(N^z_-\,+\,\imath\,M^z_-)^{-1}\phi\,\in\,\Ker(U^z-\one)$
$$
\psi^* \,\imath\,(U^z)^*\partial_z U^z\,\psi
\;=\;
2\,z^{-1}\,\|u^{\overz}_-\phi\|^2
\;.
$$
In particular, $\imath\,(U^z)^*\partial_z U^z$ restricted to the kernel of $U^z-\one$ is definite with a sign given by the sign of $z$. This implies that the path $z\mapsto \binom{N^z_-}{M^z_-}$ is transversal to $\binom{\one}{0}$ and unidirectional. Therefore the associated Maslov index is the spectral flow of $z\mapsto U^z$ through $1$ and counts exactly the eigenvalues of $H$. Let us also note that due to Proposition~\ref{prop-HighEnergy} proved below there is a neighborhood of $z=0$ with no intersections.
}
\hfill $\diamond$
\end{rem}

\section{High energy asymptotics}
\label{sec-HighEnergy}

Let us note that $z\to 0$ if and only if $|E|\to\infty$. Hence the following statement concerns the high energy asymptotics.

\begin{proposi}
\label{prop-HighEnergy} One has $\lim_{z\to 0} M^z_-=\one$. Thus $z=0$ is a removable singularity of $M^z_-$.
\end{proposi}

\noindent {\bf Proof:} Due to the factorization property \eqref{eq-TransferMProp},
$$
\Mm^z\;=\;\prod_{n=K_-}^{K_+} \Mm^z(n)
\;,
$$
where here and in the following the product is ordered with the factors ordered according to the index, smallest $n$ being on the r.h.s.. Each matrix $\Mm^z(n)$ is given by \eqref{eq-TransferMExplicit} and will be factorized as follows:
$$
{\Mm}^z(n)
\;=\;
\one
\;+\;
\imath\,\nu^z\,
\begin{pmatrix}
V(n) & 0 \\ 0 & V(n) 
\end{pmatrix}\,
\left[
\Jj
\,+\,
\begin{pmatrix}
0 & z^{-2n} \\ -z^{2n}  & 0 
\end{pmatrix}\,
\right]
\;.
$$
The matrices on the r.h.s. commute. Replacing in the above, therefore leads to
$$
\Mm^z
\;=\;
\one
\,+\,
\sum_{\emptyset\not=J\subset[K_-+1,K_+]}
(\imath\nu^z)^{|J|}
\,\left[
\prod_{n\in J}
\begin{pmatrix}
V(n) & 0 \\ 0 & V(n) 
\end{pmatrix}
\right]
\,
\prod_{n\in J}
\left[
\Jj
\,+\,
\begin{pmatrix}
0 & z^{-2n} \\ -z^{2n}  & 0 
\end{pmatrix}\,
\right]
\;,
$$
where the sum is over all subsets $J$ of $\{K_-+1,\ldots,K_+\}$ and $|J|$ denotes the cardinality of $J$. Now by \eqref{eq-MDef3} $M^z_-$ is the lower right entry of $\Mm^z$. Multiplying out the products on the r.h.s. leads to a large number of summands, but in each there will be ordered factors of the off-diagonal matrix of the type
$$
\begin{pmatrix}
0 & z^{-2n} \\ -z^{2n}  & 0 
\end{pmatrix}\,
\begin{pmatrix}
0 & z^{-2m} \\ -z^{2m}  & 0 
\end{pmatrix}\,
\;=\;
\begin{pmatrix}
-z^{2(m-n)} & 0 \\ 0 & -z^{2(n-m)}   
\end{pmatrix}
\;,
\qquad
n>m
\;,
$$
or possibly with a factor $\Jj$ between the first two factors. The crucial fact is now that the lower right entry is non-singular, as due to the ordering one has $n>m$. Therefore all lower right entries are non-singular. Moreover, $\lim_{z\to 0}\nu^z=0$. This implies the claim.
\hfill $\Box$

\section{Band edge singularities}
\label{sec-BandEdge}

At the band edges $E=\pm 2$, one has $z=\pm 1$. As $z\to \pm 1$ on the unit circle, $\nu^z\to\infty$ with a pole of first order. Due to \eqref{eq-TransferMExplicit} and \eqref{eq-TransferMProp}, this leads to singularities of the plane wave transfer matrix $\Mm^z$. On first sight, one may expect these singularities to be of order $K_+-K_-$, stemming from the multiplication of the factors $\nu^z$. However, the special structure of \eqref{eq-TransferMExplicit} implies that the singularity is only of order $1$.

\begin{proposi}
\label{prop-TransferSingularity}
There exists an anlatytic function $z\in\CM\setminus\{0\} \mapsto \Gg^z_\pm\in\CM^{2L\times 2L}$ and matrices $\Ff_\pm\in\CM^{2L\times 2L}$ such that $\Mm^z=\Gg^z_\pm+\nu^z \Ff_\pm$ in a neighborhood of $\pm 1$. 
\end{proposi}

\noindent {\bf Proof.} Let us provide two proofs. First consider the set
$$
\mathfrak{G}
\;=\;
\left\{
\begin{pmatrix}
V & z^{2n} V \\
-z^{2m} V &  -z^{2n+2m} V
\end{pmatrix}
\;:\;
V\in\CM^{L\times L}\,,\;\;n,m\in\ZM
\right\}
\;.
$$
Note that $V$ is not necessarily selfadjoint here. The set $\mathfrak{G}$ is a multiplicative subsemigroup of $\CM^{2L\times 2L}$ because for any $V,V'\in\CM^{L\times L}$ and $n,m,n',m'\in\ZM$, one has the identity
$$
\begin{pmatrix}
V & z^{2n} V \\
-z^{2m} V & \!\! -z^{2n+2m} V
\end{pmatrix}
\begin{pmatrix}
V' & z^{2n'} V' \\
-z^{2m'} V' & \!\! -z^{2n'+2m'} V'
\end{pmatrix}
=
(1-z^{2n+2m'})
\begin{pmatrix}
VV' & z^{2n'} VV' \\
-z^{2m} VV' & \!\! -z^{2n'+2m} VV'
\end{pmatrix}
\,.
$$
By \eqref{eq-TransferMExplicit} and \eqref{eq-TransferMProp},  $\Mm^z$ is in the span of $\mathfrak{G}$. Moreover, for $n+m'>0$, each such product contains a factor $(1-z^{2n+2m'})=(1-z^2)(1+z^2+\ldots+z^{2n+2m'-2})$. As $\nu^z(1-z^2)=-\imath z$, this cancels the singularity of one $\nu^z$. For $n+m'<0$, one argues similarly and for $n+m'=0$ the product vanishes. Consequently, in all products in  \eqref{eq-TransferMProp} only one singular factor $\nu^z$ remains. Extracting its singularity leads to the claim.

\vspace{.1cm}

The second proof is based on \eqref{eq-TransferMat} and an explicit calculation of the inverse of $\Cc^z$ which shows
$$
\Mm^z
\;=\;
\imath\,\nu^z\,\Dd^z(K_+)^{-1}\,
\begin{pmatrix}
-\one & z^{-1}\,\one \\ \one & -z\,\one
\end{pmatrix}
\Tt^E(K_+,K_-)
\,\Cc^z\Dd^z(K_-)
\;.
$$
Therefore one can compute the limit
$$
\mathcal{F}_{\pm}
\;=\;
\lim_{z\to \pm{1}} (\nu^z)^{-1}\mathcal{M}^z
\;=\;
\imath\begin{pmatrix}-\one & \pm\,\one \\ \one & \mp\,\one \end{pmatrix}
\mathcal{T}^2(K_+,K_-)
\begin{pmatrix}
\pm\one & \pm\,\one \\	 \one & \,\one\end{pmatrix}
\;,
$$ 
and deduce $\mathcal{G}^z_{\pm}=\mathcal{M}^z-\nu^z\mathcal{F}_{\pm}$.
\hfill $\Box$

\vspace{.2cm}

The matrices $\Ff_\pm$ in Proposition~\ref{prop-TransferSingularity} can also be expressed in terms of Wronskians of Jost solutions. Indeed, comparing with \eqref{eq-MDef2}, one finds 
$$
\Ff_\pm
\;=\;
\lim_{z\to \pm 1}
\,
(\nu^z)^{-1}\Mm^z
\;=\;
\lim_{z\to \pm 1}
\begin{pmatrix}
W(u^{\overzb}_+,u^z_-)   & W(u^{\overzb}_+,u^{\overz}_-)  
\\
-\,W(u^{\zb}_+,u^z_- )  &  -\,W(u^{\zb}_+,u^{\overz}_-)  
\end{pmatrix}
\;.
$$
Furthermore Proposition~\ref{prop-TransferSingularity} implies that $M^z_-$ as a matrix entry of $\Mm^z$ has a similar singularity behavior, namely in a neighborhood of $\pm 1$
$$
M^z_-
\;=\;G^z_\pm\,+\,\nu^z\,F_\pm\;,
$$
for trigonometric polynomials $G^z_\pm\in\CM^{L\times L}$ and matrices $F_\pm\in\CM^{L\times L}$. Please note again that the subscripts $\pm$ correspond to the upper/lower band edge at $\pm 2$ and are not related to the subscript on $M^z_-$. Comparing with the above,  one deduces 
\begin{equation}
\label{F_+}
F_\pm 
\;=\;
\lim_{z\to \pm 1}-\,W(u^{\zb}_+,u^{\overz}_-)
\;=\;-\,W(u_+^1,u_-^1)
\;.
\end{equation} 
For further analysis, let us focus on $F=F_+$ at the upper band edge. The lower band edge is similar. The singular value decomposition is $F=UDU'$ where $U$ and $U'$ are unitary and $D\geq 0$ is diagonal. If $F$ is not singular so that $D>0$, one has
$$
T^1_-
\;=\;
\lim_{z\to 1} \,\big(G^z\,+\,\nu^z\,F\big)^{-1}
\;=\;
\lim_{z\to 1}\, (U')^*\big(U^*G^z(U')^*\,+\,\nu^z\,D\big)^{-1}U^*
\;=\;0
\;.
$$
Also the coefficient matrix $N^z_-$ has a decomposition $N^z_-=\hat{G}^z+\nu^z \hat{F}$ with
$$
\hat{F}
\;=\;
\lim_{z\to 1}\,W(u^{\overzb}_+,u^{\overz}_-)
\;=\;
W(u^1_+,u_-^1)
\;.
$$ 
Consequently one finds in the non-singular case
\begin{equation}\label{R^1_}
R^1_-
\;=\;
-\,\lim_{z\to 1} \,N^z_-(M_-^z)^{-1}
\;=\;
-\,\lim_{z\to 1} \,(\hat{G}^z+\nu^z \hat{F})(G^z+\nu^z F)^{-1}
\;=\;
-\,\hat{F}\,F^{-1}
\;=\;
\one
\;.
\end{equation}

Next let us turn to the case of a singular matrix $F$. Then $D=\diag(0,f)$ with $f>0$. For later use, let us set $J_h^1=L-$rank$(f)=L-$rank$(F)=\dim\,\Ker(F)$. Then
$$
T^z_-
\;=\;
\left(
G^z +\nu^z
UDU'
\right)^{-1}
\;=\;
(U')^*
\left(
U^*G^z(U')^* +\nu^z
\diag(0,f)
\right)^{-1}
U^*
\;.
$$
Now the inverse can be calculated with the Schur complement formula (for the grading of $D$). Set
$$
U^*G^z(U')^*
\;=\;
\begin{pmatrix} a^z & b^z \\ c^z & d^z
\end{pmatrix}
\;.
$$
Then
\begin{equation}
\label{eq-UTU}
U'T^z_-U
\;=\;
\begin{pmatrix}
(s^z)^{-1} & -(s^z)^{-1}b^z(d^z+\nu^zf)^{-1} \\
-(d^z+\nu^zf)^{-1}c^z(s^z)^{-1} &  \ \ (d^z+\nu^zf)^{-1}+ (d^z+\nu^zf)^{-1}c^z(s^z)^{-1}b^z(d^z+\nu^zf)^{-1}
\end{pmatrix}
\;,
\end{equation}
with Schur complement $s^z=a^z-b^z(d^z+\nu^zf)^{-1}c^z$. Recall the general fact the Schur complement of a matrix with invertible lower right entry is invertible if and only if the matrix itself is invertible. Here the matrix $U'T^z_-U=U'(M^z_-)^{-1}U$ is invertible for $z\in\SM^1_0=\SM^1\setminus\{-1,1\}$. Moreover, $\|U'T^z_-U\|=\|T^z_-\|=\|(M^z_-)^{-1}\|\leq 1$ for $z\in\SM^1_0$ due to $(M^z_-)^*M^z_-\geq \one$ (see the proof of Lemma~\ref{lem-Minvertible}). Therefore also $\|(s^z)^{-1}\|\leq 1$ for $z\in\SM^1_0$. On the other hand, $f>0$ implies $\lim_{z\to 1}s^z=a^1$ so that $a^1$ is invertible. It follows
$$
T^1_-
\;=\;
(U')^*
\begin{pmatrix}
(a^1)^{-1} & 0 \\ 0 & 0 
\end{pmatrix}
U^*
\;.
$$
Furthermore, one has $R^z_-=(\hat{G}^z+\nu^z \hat{F})T^z_-$. As the limit $T^1_-$ is bounded and $\|R^z_-\|\leq 1$, one must have $\hat{F}T^1_-=0$ so that 
$$
R^1_-\;=\; \hat{G}^1 T^1_-
\;.
$$
Formulas for $T_+^1$ and $R_+^1$ can be obtained in a similar manner. Another set of formulas for the limits result from the identities \eqref{eq-MNId0}, notably
$$
T^1_+
\;=\;
\lim_{z\to 1}
\;\imath(z^{-1}-z) \;W(u^{\overzb}_-,u^z_+)^{-1}
\;,
\qquad
T^1_-
\;=\;
\lim_{z\to 1}
\;\imath(z-z^{-1}) \;W(u^{\zb}_+,u^{\overz}_-)^{-1}
\;,
$$
and similarly for $R^1_\pm$. This implies
$$
T^1_-
\;=\;
\lim_{z\to 1}
\;\imath(\zb-\zb^{-1}) \;W(u^{z}_+,u^{\overzb}_-)^{-1}
\;=\;
\Big(\lim_{z\to 1}
\;\imath(z^{-1}-z) \;W(u^{\overzb}_-,u^z_+)^{-1}\Big)^*
\;=\;
(T^1_+)^*
\;.
$$
Similarly, one checks $R^1_-=(R^1_+)^*$. Analogous formulas for $z=-1$ can be written out. Due to the identity $(M^z_-)^{-1}=T^z_-$, this implies the following fact that will be used later on.

\begin{proposi}
\label{prop-MInverse}
The limits $\lim_{z\to\pm 1}(M^z_-)^{-1}$ exist. 
\end{proposi}

\section{Half-bound states}
\label{sec-AsympConst}

At the band edge $E=2$ (and similarly $E=-2$), one has $z= 1$. In the limit $z\to 1$, the solutions $u_+^{\bar{z}}$ and $u_-^{1/z}$ can possibly aline and then their Wronskian $W(u_+^{\bar{z}},u_-^{1/z})$ converges to $0$, at least on a subspace. Hence its inverse will diverge there. In the last section, it was shown that the associated pole can be extracted and determines the limit of the scattering matrix. The non-generic behavior where \eqref{eq-Generic} does not hold, is connected to the existence of special states at band edge energies. Indeed, if the Wronskian $W(u_+^{\bar{z}},u_-^{1/z})$ has a kernel in the limit $z\to 1$, then the alignment of directions of $u_+^{\bar{z}}$ and $u_-^{1/z}$ in the limit $z\to 1$ allows to construct bounded solutions, lying in $\ell^\infty(\ZM,\CM^L)$ and thus the span of  the bounded (Jost) solutions $u^1_\pm$ at $\pm\infty$.  Similar as in \eqref{eq-FrameDef0} they lead to matrices 
$$
\Phi^1_\pm(n)
\;=\;
\begin{pmatrix}
\one \\ \one
\end{pmatrix}
\;\in\;\CM^{2L\times L}
\;,
\qquad
n>K_+\;\mbox{ or }\;n<K_-\;\mbox{ respectively }.
$$
These matrices satisfy $(\Phi^1_\pm)^*\Ii \Phi^1_\pm=0$ and are thus $\Ii$-Lagrangian. Asymptotically constant solutions (both at $+\infty$ and $-\infty$) are constructed from vectors in 
\begin{equation}
\label{eq-Intersec}
\Ran\big(\Phi^1_+(K_+)\big)\cap \Ran\big(\Tt^2(K_+,K_-)\Phi^1_-(K_-)\big)
\;,
\end{equation}
provided the intersection is non-trivial. Then choosing $\phi\in\CM^L$  such that $\Phi^1_+(K_+)\phi$ is in the intersection, one then has a bounded solution
$$
u_\phi(n)\;=\;\binom{0}{\one}^*\Tt^2(n,K_+)\Phi^1_+(K_+)\phi
\;.
$$
Such solutions are constant outside of $\{K_-+1,\ldots,K_+\}$. In the case of continuous Schr\"odinger operators, such solutions are called half-bound states because they have a contribution $\frac{1}{2}$ in Levinson's theorem. These states are not in the Hilbert space, but may, depending on dimension, still decay \cite{JK}. Here theses solutions are asymptotically constant. Of course, these half-bound states do not lead to eigenvalues of $H$ as a selfadjoint operator. 

\vspace{.2cm}

Now let $J^+_h$ be the dimension of this intersection. By construction, it is equal to the dimension of the space of bounded solutions at energy $E=2$ and thus $z=1$ (that is, the dimension of half-bound states at $z=1$). There is a similar dimension $J^{-}_h$ for $E=-2$ and $z=-1$. 

\begin{proposi}
\label{prop-TransferSingularityDim}
One has $J_h^\pm=\dim\,\Ker(F_\pm)$ with $F_\pm$ as defined in {\rm Section~\ref{sec-BandEdge}} at $z=\pm 1$. Moreover, the map $z\mapsto \det(M^z_-)$ has a pole of order $L-J_h^\pm$ at $z=\pm 1$.
\end{proposi}

\noindent {\bf Proof.} 
Let us focus on $z=1$. By \eqref{F_+},
$$
F_+
\;=\;
-\,W(u_+^1,u_-^1)
\;=\;
\imath(u^1_-(n+1)-u^1_-(n))
\;,
$$
for all $n\geq K_+$. It follows that  
$$
\lim_{n\to\infty}\,\frac{u^1_-(n)}{n} 
\;=\; 
\lim_{n\to\infty}\,\frac{\sum_{m=0}^{n-1}{(u^1_-(m+1)-u^1_-(m)})}{n}
\;=\;
-\,\imath\, F_+ 
\;.
$$
As $(u_+^1,v_+^1)$ is a fundamental solution at $E=1$, there are $L\times L$ matrices $X,Y$ such that $u_-^1=u_+^1X+v_+^1Y$. Using that $u^1_+(n)=\mathbf{1}$ and $v_+^1(n)=n$ for $n\geq K_+$, it follows that $Y=-\imath F_+$. Therefore, $u_-^1(n)=X -\imath nF_+$ for $n \geq K_+$. Therefore solutions of the form $u_-^1\phi$ are bounded if and only if $\phi \in \Ker(F_+)$. This implies that $J_h^+=\dim(\Ker(F_+))$.

\vspace{.1cm}

As to the second claim, it is equivalent to $z\mapsto \det((M^z_-)^{-1})=\det(T^z_-)$ having a zero of order $L-J_h^\pm$ at $z=\pm 1$. This will be verified for $z=1$ using the explicit formulas for $T^z_-$ given in Section~\ref{sec-BandEdge}. The case $z=-1$ is analogous. For $U,U'$ as given there, let us introduce the following notation for the matrix entries of  $U'T^z_-U$ as given in \eqref{eq-UTU}:
$$
U'T^z_-U
\;=\;
\begin{pmatrix}
\hat{a}^z & \hat{b}^z \\
\hat{c}^z & \hat{d}^z
\end{pmatrix}
\;,
$$
namely $\hat{a}^z=(s^z)^{-1}$, $\hat{b}^z=-(s^z)^{-1}b^z(d^z+\nu^zf)^{-1}$ and so on. As $f>0$ and $\|d^z\|\leq 1$ for $z\in\SM^1_0$, one has 
$$
\lim_{z\to 1}
\,(d^z+\nu^zf)^{-1} 
\;=\; 
0\;,
\qquad
\lim_{z\to 1}
\,\nu^z(d^z+\nu^zf)^{-1} 
\;=\; 
f^{-1}
\;.
$$
This leads to
$$
\lim_{z\to 1}
\;\hat{b}^z(\hat{d}^z)^{-1}\hat{c}^z
\;=\;
\lim_{z\to 1}
\;(s^z)^{-1}b^z(d^z+\nu^zf)^{-1}
\big(\one+ c^z(s^z)^{-1}b^z(d^z+\nu^zf)^{-1}\big)^{-1}
c^z(s^z)^{-1}
\;=\;
0\;.
$$
Notice that $\hat{d}^z$ is invertible because $\nu^z\hat{d}^z \to f^{-1}$ by \eqref{eq-UTU}. Hence 
$$
\lim_{z\to 1}
\;
\det\big(\hat{a}^z-\hat{b}^z{(\hat{d}^z)}^{-1}\hat{c}^z\big) 
\;=\; 
\lim_{z\to 1}
\;
\det
\big((s^z)^{-1}\big)
\;=\;
\det\big((a^1)^{-1}\big)
\;\neq 0
\;,
$$
because $a^1$ is invertible. Therefore $\hat{a}^z-\hat{b}^z{(\hat{d}^z)}^{-1}\hat{c}^z$ is invertible in a neighborhood of $1$. Similarly, 
$$
\lim_{z\to 1}
\det\big(\nu^z\hat{d}^z\big)
\;=\;
\det(f^{-1})
\;.
$$
Finally, using Schur formula for the determinant,
$$
\det(U'T^z_-U)
\;=\;
(\nu^z)^{-L+J_h^+}\,
\det\big(\nu^z\hat{d}^z\big)
\,
\det\big(\hat{a}^z-\hat{b}^z{(\hat{d}^z)}^{-1}\hat{c}^z\big)
\;,
$$
because the size and rank of $f$ is $L-J_h^+$. Due to the asymptotics stated above and the fact that $\nu^z$ has a pole of order $1$, this implies the result.
\hfill $\Box$

\begin{rem}
{\rm
Let us provide another proof of $J_h^\pm=\dim\,\Ker(F_\pm)$. It is of geometric nature and similar to the one in Remark~\ref{sec-GeoArg}. The intersection \eqref{eq-Intersec} is again the intersection of two $\Ii$-Lagrangian planes so that its dimension $J^+_h$ can be calculated by intersection theory (Proposition~2 in \cite{SB2}) as  
$$
J^+_h
\;=\;
\dim\;\Ker
\left(
\Phi^1_+(K_+)^*\, \Ii\,\Tt^2(K_+,K_-)\Phi^1_-(K_-)
\right)
\;.
$$
But as
$$
\Phi^1_+(K_+)^*\;=\;
\lim_{z\to 1} \,(\imath\,\nu^z)^{-1}\,\binom{0}{\one}^*\,(\Cc^z)^{-1}\,\Ii^*
\;,
\qquad
\Phi^1_-(K_-)\;=\;
\lim_{z\to 1} \,\Cc^z\,\binom{0}{\one}
\;,
$$
one has due to \eqref{eq-TransferMat}
\begin{align*}
\Phi^1_+(K_+)^*\, \Ii\,\Tt^2(K_+,K_-)\Phi^1_-(K_-)
&
\;=\;
\lim_{z\to 1}
\;(\imath \nu^z)^{-1}\,\binom{0}{\one}^*
\Mm^z\,\binom{0}{\one}\,
\\
&
\;=\;
\lim_{z\to 1}
\,(\imath\nu^z)^{-1}\,M^z_-
\\
&
\;=\;
-\,\imath\,F\;,
\end{align*}
where $F=F^+$ is as in Section~\ref{sec-BandEdge}, which then also implies $J^+_h=\dim\, \Ker(F)$.
}
\hfill $\diamond$
\end{rem}

Let us also collect the above information on the half-bound states in the following result.

\begin{proposi}
\label{prop-AsympConst}
For each $\phi\in\Ran(T_+^{ 1})$, there exists a state $u_\phi\in\ell^\infty(\ZM,\CM^L)$ satisfying the Schr\"odinger equation $Hu_\phi= 2u_\phi$. The limits
$$
\lim_{n\to\pm \infty}\,u_\phi(n)
$$
exist. The dimension $J^+_h$ of the space of these half-bound states is equal to $\dim(\Ran(T_+^{ 1}))$. A similar statement hold for $z=-1$.
\end{proposi}

\section{Time delay}
\label{sec-TimeDelay}

Levinson's theorem concerns the winding of the scattering matrix on the unit circle. The integrand is also called the time delay. It can then be calculated by the following general principle related merely to the passage \eqref{eq-VM} from the $\Jj$-unitary $\Mm$ to the unitary $\Vv(\Mm)$.

\begin{proposi}
\label{prop-VDeriv}
Let $t\mapsto \Mm_t$ be a differentiable path of $\Jj$-unitaries with diagonal entries $A_t$ and $D_t$. Then
$$
\Tr\big(\Vv(\Mm_t)^*\partial_t\Vv(\Mm_t)\big)
\;=\;
\Tr\big((A_t)^{-1}\partial_t A_t\,-\,(D_t)^{-1}\partial_t D_t\big)
\;.
$$
\end{proposi}

\noindent {\bf Proof.} Let us drop the index $t$ and also simply write $\partial=\partial_t$. The $\Jj$-unitarity of $\Mm$ and $\Mm^*$ is equivalent to the following identities 
implies
\begin{align*}
A^*A\;=\;\one+C^*C\;,
\qquad
D^*D\;=\;\one+B^*B\;,
\qquad
A^*B\;=\;C^*D
\;,
\\
AA^*\;=\;\one+BB^*\;,
\qquad
DD^*\;=\;\one+CC^*\;,
\qquad
AC^*\;=\;BD^*
\;.
\end{align*}
As already noted, $A$ and $D$ are thus invertible. Now
\begin{align*}
\Tr\big(\Vv(\Mm)^*\partial\Vv(\Mm)\big)
&
\;=\;
\Tr
\left(
\begin{pmatrix}
A^{-1} & A^{-1}B \\
-(D^*)^{-1}B^* & (D^*)^{-1}
\end{pmatrix}
\partial
\begin{pmatrix}
(A^*)^{-1} & -BD^{-1} \\
B^*(A^*)^{-1} & D^{-1}
\end{pmatrix}
\right)
\\
&
\;=\;
\Tr
\big(
A^{-1}\partial (A^*)^{-1} 
+
A^{-1}B\partial B^*(A^*)^{-1}
+
A^{-1}B B^*\partial(A^*)^{-1}
\\
&
\;\;\;\;\;\;\;\;\;\;\;\;
+
(D^*)^{-1}B^*  \partial B D^{-1}
+
(D^*)^{-1}B^* B \partial D^{-1}
+
(D^*)^{-1}\partial D^{-1}
\big)
\;.
\end{align*}
Now let us replace $BB^*$ and $B^*B$ by the above expressions in the third and fifth summand\textcolor{blue}{s}:
\begin{align*}
\Tr\big(\Vv(\Mm)^*\partial\Vv(\Mm)\big)
&
\;=\;
\Tr
\big(
A^*\partial (A^*)^{-1} 
+
A^{-1}B\partial B^*(A^*)^{-1}
+
(D^*)^{-1}B^*  \partial B D^{-1}
+
D\partial D^{-1}
\big)
\\
&
\;=\;
\Tr
\big(
A^*\partial (A^*)^{-1} 
+
(A^*)^{-1}A^{-1}B\partial B^*
+
D^{-1}(D^*)^{-1}B^*  \partial B 
+
D\partial D^{-1}
\big)
\\
&
\;=\;
\Tr
\big(
A^*\partial (A^*)^{-1} 
+
(AA^*)^{-1}B\partial B^*
+
(D^*D)^{-1}B^*  \partial B 
+
D\partial D^{-1}
\big)
\;.
\end{align*}
Now replace $AA^*$ and $D^*D$ in terms of $B$ and use $(\one +B^*B)^{-1}B^*=B^*(\one+BB^*)^{-1}$. Again using the cyclicity one finds
\begin{align*}
\Tr\big(\Vv(\Mm)^*\partial\Vv(\Mm)\big)
&
\;=\;
\Tr
\big(
A^*\partial (A^*)^{-1} 
+
(\one+BB^*)^{-1}\partial (BB^*)
+
D\partial D^{-1}
\big)
\\
&
\;=\;
\Tr
\big(
A^*\partial (A^*)^{-1} 
+
(AA^*)^{-1}\partial (AA^*)
+
D\partial D^{-1}
\big)
\\
&
\;=\;
\Tr
\big(
-(A^*)^{-1}\partial A^* 
+
(AA^*)^{-1}(\partial AA^*+A\partial A^*)
-
D^{-1}\partial D
\big)
\;,
\end{align*}
which implies the result.
\hfill $\Box$

\vspace{.2cm}

When applied to \eqref{eq-VScat}, one gets a formula for the time delay for $z\in\SM^1\setminus\{-1,1\}$. As both sides are meromorophic, this formula extends to all $\{z\in \mathbb{C} : z, z^{-1} \in \mathbb{C}_0\}$.

\begin{coro}
\label{coro-TimeDelay}
For all $z,\overline{z}^{-1}\in\CM_0\cup\{-1,1\}$,
\begin{align}
\label{eq-VScatDer}
\Tr
\big(
(\Ss^{\overzb})^*\partial_z\Ss^z
\big)
& 
\;=\;
\Tr
\big(
(M^{\overz}_-)^{-1}\partial_z M^{\overz}_-
-
(M_-^z)^{-1}\partial_z M^z_-
\big)
\\
&
\;=\;
\det (M^{\overz}_-)^{-1}\,\partial_z\,\det (M^{\overz}_-)\;-\;
\det (M^{z}_-)^{-1}\,\partial_z\,\det (M^{z}_-)
\;.
\end{align}
%
\end{coro}

\section{Levinson-type theorem}
\label{sec-Levinson}

This section is only devoted to the

\vspace{.2cm}

\noindent {\bf Proof of Theorem~\ref{theo-Levinson}.} Let us first prove the statements about the spectrum of $H$. 
Due to Proposition~\ref{prop-BoundStates} the eigenvalues are the zeros of $z\mapsto\det(M^z_-)$ and by Proposition~\ref{prop-MInverse} there cannot be any accumulation of eigenvalues at $-2$ and $2$. As the operator $H$ is bounded, there are indeed only a finite number of eigenvalues outside of $[-2,2]$. Furthermore, embedded eigenvalues lying in $[-2,2]$ will have have oscillating eigenfunctions outside of the support of $V$ so that they actually have to vanish outside of this support. However, compactly supported eigenfuntions vanish identically due to the three-term recurrence equation resulting from \eqref{eq-Hamiltonian}.

\vspace{.1cm}

Let now $\SM^1_{\pm\epsilon}$ be the the positively oriented circle of radius $1\pm\epsilon$. Let us integrate \eqref{eq-VScatDer} over this path. Then
$$
\oint_{\SM^1_{-\epsilon}}\frac{dz}{2\pi\imath}\;
\Tr
\big(
(\Ss^{\overzb})^*\partial_z\Ss^z
\big)
\;=\;
-\,
\left(\oint_{\SM^1_{\epsilon}}
+
\oint_{\SM^1_{-\epsilon}}
\right)
\frac{dz}{2\pi\imath}\;
\det (M^{z}_-)^{-1} \partial_z\,\det (M^{z}_-)
\;.
$$
Now one can invoke the argument principle.  According to Proposition~\ref{prop-BoundStates}, Corollary~\ref{coro-C0} and Proposition~\ref{prop-TransferSingularityDim}, the map $z\mapsto \det(M^z_-)$ is meromorphic on a neighborhood of the unit disc with zeros at $z$ such that $z+z^{-1}$ is an eigenvalue both counted with their multiplicity and with poles of order $L-J_h^\pm$ at $z=\pm 1$, and no other zeros or poles. Therefore the integrals can be evaluated: 
$$
\oint_{\SM^1_{-\epsilon}}\frac{dz}{2\pi\imath}\;
\Tr
\big(
(\Ss^{\overzb})^*\partial_z\Ss^z
\big)
\;=\;
L\,-\,J_h^+\,-\,J_b\,+L\,-\,J_h^-\,-\,J_b
\;,
$$
for $\epsilon$ sufficiently small.  Now the integrand on the l.h.s. is analytic in a neighborhood of the $\SM^1$ (see Theorem \ref{theo-ScatAnaly}) so that one can take the limit $\epsilon\to 0$. As on $\overzb=z$ on $\SM^1$,
$$
\oint_{\SM^1}\frac{dz}{2\pi\imath}\;
\Tr
\big(
(\Ss^z)^*\partial_z\Ss^z
\big)
\;=\;
2\,L\,-\,J_h\,-\,2\,J_b
\;,
$$
where $J_h=J_h^++J_h^-$ is the total number of half-bound states. Finally let us split $\SM^1$ into upper and lower arcs $\SM^1_+$ and $\SM^1_-$, both with positive orientation. Parametrizing the upper arc with $k\in[0,\pi]\mapsto e^{\imath k}\in\SM^1$ followed by the change variables $k\in[0,\pi]\mapsto E=2\cos(k)$ shows
$$
\oint_{\SM^1_+}\frac{dz}{2\pi\imath}\;
\Tr
\big(
(\Ss^z)^*\partial_z\Ss^z
\big)
\;=\;
\int^\pi_0
\frac{dk}{2\pi\imath}\;
\Tr
\big(
(\Ss^{e^{\imath k}})^*\partial_k\Ss^{e^{\imath k}}
\big)
\;=\;
\int^2_{-2}
\frac{dE}{2\pi\imath}\;
\Tr
\big(
(\Ss^E)^*\partial_E\Ss^E
\big)
\;,
$$
as according to our convention $\Ss^E=\Ss^{e^{\imath k}}$. 
As to the lower circle, let us use the parametrization $[-\pi,0]\mapsto e^{\imath k}$ and the identities $\Kk\Ss^{e^{\imath k}}\Kk=(\Ss^{e^{-\imath k}})^*$ and $\Kk\partial_k\Ss^{e^{\imath k}}\Kk=(\partial_k\Ss^{e^{-\imath k}})^*$, so that due to the cyclicity of the trace and $\Kk^2=\one$:
\begin{align*}
\oint_{\SM^1_-}\frac{dz}{2\pi\imath}\;
\Tr
\big(
(\Ss^z)^*\partial_z\Ss^z
\big)
&\;=\;
\int^0_{-\pi}
\frac{dk}{2\pi\imath}\;
\Tr
\big(
(\Ss^{e^{\imath k}})^*\partial_k\Ss^{e^{\imath k}}
\big)
\\
&
\;=\;
\int^0_{-\pi}
\frac{dk}{2\pi\imath}\;
\Tr
\big(
\Ss^{e^{-\imath k}}\partial_k (\Ss^{e^{-\imath k}})^*
\big)
\\
&
\;=\;
-\int_{-\pi}^{0}
\frac{dk}{2\pi\imath}\;\Tr((\mathcal{S}^{e^{-\imath k}})^*\partial_k\mathcal{S}^{e^{-ik}})
\\
&\;=\;
\int_{0}^{\pi}\frac{dk}{2\pi\imath}\;\Tr((\mathcal{S}^{e^{\imath k}})^*\partial_k\mathcal{S}^{e^{ik}})
\\
&\;=\;
\int_{\mathbb{S}^1_+}\frac{dz}{2\pi\imath}\;\Tr((\mathcal{S}^z)^*\partial_z\mathcal{S}^z)
\;,
\end{align*}
where in the third equality results from $\Tr((\mathcal{S}^{e^{-\imath k}})^*\partial_k\mathcal{S}^{e^{-\imath k}})=-\Tr(\partial_k(\mathcal{S}^{e^{-\imath k}})^*\mathcal{S}^{e^{-\imath k}})$, which in turn follows from Proposition \ref{prop-TransScatProp}(iii) for $\overz^{-1}=z\in\SM^1$. Therefore the contribution of $\SM^1_-$ is the same as that of $\SM^1_+$ and this concludes the proof.
\hfill $\Box$



\section{Green function and scattering matrix}
\label{sec-GreenFunction}

Recall the definition \eqref{eq-GreenDef} of the Green function. Note that $\Im m(G^E(n,n))>0$ for $\Im m(E)>0$, which is hence invertible. Furthermore, the function $z\mapsto G^{z+z^{-1}}(n,m)$ is analytic on $\CM\setminus\SM^1$.  The following can readily be deduced from \cite{SB0}, but for sake of completeness we provide a complete proof.

\begin{proposi}
\label{prop-Green}
For $E=z+\overzz\not\in\sigma(H)$ with $|z|<1$,
\begin{align*}
&
M^z_-
\;=\;
\frac{z^{K_+-K_-}}{z-z^{-1}}
\;
G^E(K_-,K_+)^{-1}
\;,
\\
&
N^z_-
\;=\;
z^{-K_+-K_-}\,
G^E(K_+,K_+)\,
G^E(K_-,K_+)^{-1}
\;-\;
\frac{z^{-K_+-K_-}}{z-z^{-1}}
\;
G^E(K_-,K_+)^{-1}
\;.
\end{align*}
\end{proposi}

\noindent {\bf Proof.} 
Let us set $G^E(n)=G^E(n,K_+)$ and view it as a matrix-valued function on $\ZM$. Then
%
%
%
$$
G^E(n+1)\,+\,G^E(n-1)\,+(V(n)-E)G^E(n)
\;=\;
\delta_{n,K_+}\,\one_L
\;.
$$
Let us introduce
$$
\Psi^E(n)
\;=\;
\begin{pmatrix}
G^E(n+1) \\ G^E(n)
\end{pmatrix}
\;.
$$
Using the three term recurrence relation iteratively, one sees that $\Psi^E(n)$ is of full rank $L$. Furthermore, one has
$$
\Psi^E(n)
\;=\;
\Tt^E(n)
\Psi^E(n-1)
\;+\;
\delta_{n,K_+}\,\binom{\one}{0}
\;.
$$
In particular,
\begin{equation}
\label{eq-ConnectSides}
\Psi^E(K_+)
\;=\;
\Tt^E(K_+,K_-)\,\Psi^E(K_-)
\;+\;\binom{\one}{0}
\;.
\end{equation}
As the matrix $\Psi^E(n)$ is expressed in terms of the resolvent, it decays both as $n\to\infty$ and $n\to-\infty$ because $n\mapsto G^E(n)$ is square summable. For $|z|<1$, it follows from the text below \eqref{eq-PhiBuild} and similar arguments that there are solutions $\hat{\Psi}_+^E$ and $\hat{\Psi}_-^E$ with initial conditions $\Psi^E(K_+)$ and $\Psi^E(K_-)$ that decay at $+\infty$ and $-\infty$, respectively, in a square summable fashion. This implies that  there are square matrices $\alpha^z_\pm$ such that  $\hat{\Psi}_+^E=\Phi^z_+ \alpha^z_+$ and $\hat{\Psi}_-^E=\Phi^{\overz}_-\alpha^z_-$. In particular,
$$
\Psi^E(K_+)
\;=\;
\Phi^z_+(K_+)\,\alpha^z_+
\;,
\qquad
\Psi^E(K_+)
\;=\;
\Phi^{\overz}_-(K_+)\,\alpha^z_-
\;.
$$
Because $\Psi^E(K_\pm)$ are of full rank $L$, the matrices $\alpha^z_\pm$ are invertible. From the lower equations of these identities, one deduces 
$$
\alpha^z_+
\;=\;
z^{-K_+}\,
G^E(K_+)
\;,
\qquad
\alpha^z_-
\;=\;
z^{K_-}\,
G^E(K_-)
\;.
$$
Moreover,
\begin{align*}
\Phi^{\overz}_-(K_+)
&
\;=\;
\Tt^E(K_+,K_-)\,\Phi^{\overz}_-(K_-)
\;=\;
\Tt^E(K_+,K_-)\,\Psi^E(K_-)\,(\alpha^z_-)^{-1}
\\
&
\;=\;
\left[
\Psi^E(K_+)\,-\,\binom{\one}{0}\right]\,(\alpha^z_-)^{-1}
\;=\;
\left[
\Phi^z_+(K_+)\alpha^z_+\,-\,\binom{\one}{0}\right]\,(\alpha^z_-)^{-1}
\;.
\end{align*}
Now using equations \eqref{eq-Wronskians} and \eqref{eq-MNId0}, one deduces
\begin{align*}
M^z_-
&
\;=\;
-\,\nu^z\;\Phi^{\zb}_+(K_+)^*\,\tfrac{1}{\imath}\,\Ii\,\Phi^{\overz}_-(K_+)
\\
&
\;=\;
-\,\nu^z\;\Phi^{\zb}_+(K_+)^*\,\tfrac{1}{\imath}\,\Ii\,\left[
\Phi^z_+(K_+)\alpha^z_+\,-\,\binom{\one}{0}\right]\,(\alpha^z_-)^{-1}
\\
&
\;=\;
0\;+\;\nu^z\;\tfrac{1}{\imath}\;\Phi^{\zb}_+(K_+)^*\,\binom{0}{\one}\,(\alpha^z_-)^{-1}
\\
&
\;=\;
\frac{z^{K_+}}{z-z^{-1}}\;(\alpha^z_-)^{-1}
\;=\;
\frac{z^{K_+-K_-}}{z-z^{-1}}
\;
G^E(K_-,K_+)^{-1}
\;.
\end{align*}
Similarly
\begin{align*}
N^z_-
&
\;=\;
\nu^z\;\Phi^{\overzb}_+(K_+)^*\,\tfrac{1}{\imath}\,\Ii\,\Phi^{\overz}_-(K_+)
\\
&
\;=\;
\nu^z\;\Phi^{\overzb}_+(K_+)^*\,\tfrac{1}{\imath}\,\Ii\,\left[
\Phi^z_+(K_+)\alpha^z_+\,-\,\binom{\one}{0}\right]\,(\alpha^z_-)^{-1}
\\
&
\;=\;
\alpha^z_+(\alpha^z_-)^{-1}\;-\;\nu^z\;\tfrac{1}{\imath}\;\Phi^{\overzb}_+(K_+)^*\,\binom{0}{\one}\,(\alpha^z_-)^{-1}
\\
&
\;=\;
z^{-K_+-K_-}\,
G^E(K_+,K_+)\,
G^E(K_-,K_+)^{-1}
\;-\;
\frac{z^{-K_+-K_-}}{z-z^{-1}}
\;
G^E(K_-,K_+)^{-1}
\;.
\end{align*}
This shows the two identities.
\hfill $\Box$

\vspace{.2cm}

Let us note that Proposition~\ref{prop-Green} also provides an alternative proof of Proposition~\ref{prop-MComplexInvert}. Furthermore, it implies a formula for the transmission matrix $T^z_-$ and reflection matrix $R^z_-$ in terms of the Green matrix:
\begin{equation}
\label{T^z_-}
T^z_-
\;=\;
z^{K_--K_+} (z-z^{-1})
\;
G^E(K_-,K_+)
\;,
\quad
R^z_-
\;=\;
z^{-2K_+}\big(-\one\;+\;(z-z^{-1})\,G^E(K_+,K_+)\big)
\;,
\end{equation}
where $|z|<1$ and $z+z^{-1}\notin \sigma(H)$. The other coefficients of the scattering matrix for $|z|<1$ can be obtained by using the relation
$$
G^{\overline{E}}(n,m)^*\;=\;G^E(m,n)
\;.
$$
Due to $T_+^z=(T_-^{\bar{z}})^*$ and the unitarity requirement, one finds
\begin{equation}
\label{T^z_+}
T^z_+
\;=\;
z^{K_--K_+} (z-z^{-1})
\;
G^E(K_+,K_-)
\;,
\quad
R^z_+
\;=\;
z^{2K_-}\big(\one\;-\;(z-z^{-1})\,G^E(K_-,K_-)\big)
\;.
\end{equation}
Hence the scattering matrix $\Ss^z$ for $|z|<1$ can be compactly written as stated in Proposition~\ref{prop-GreenScat}, thus completing the proof of this result.

\section{Examples}

If $K_+=K_-$ so that $V=0$, then there is no obstacle and the Jost solutions satisfy $u^z_+=u^z_-$. Then the transfer matrix as well as the scattering matrix is the identity. Next let us consider the case of a perturbation on one site, say site $1$. Thus $K_+=1$ and $K_-=0$. Then one deals only with $\Mm^z(1,0)=\Mm^z(1)$ given by \eqref{eq-TransferMExplicit} and, using the equations \eqref{eq-MDef3} and \eqref{eq-ScatMN2}, one finds
$$
{\Ss}^z
\;=\;
\begin{pmatrix}
\big(\one-\imath\,\nu^zV(1)\big)^{-1} & 
-\imath\,\nu^zz^{-2} V(1) \big(\one-\imath\,\nu^z V(1)\big)^{-1}
\\ -\imath\,\nu^z z^2V(1) \big(\one-\imath\,\nu^zV(1)\big)^{-1} & 
\big(\one-\imath\,\nu^zV(1)\big)^{-1}
\end{pmatrix}
\;.
$$
Note that if $|z|=1$, the inverse indeed always exists because the real part of $\one-\imath\,\nu^zV(1)$ is positive. Further, if $V(1)$ is of full rank, one obtains \eqref{eq-Generic} in the limit $z\to  1$ because then $\nu^z\to\infty$. In general, let $P$ be the projection onto the kernel of $V(1)$, then $T^1_+=P$. The asymptotically constant solution $u_\phi$ of Proposition~\ref{prop-AsympConst} can be chosen constant in this case. Therefore $V$ has a reflectionless channel if $P$ is non-trivial.

\vspace{.1cm}

Next let $K_+=1$ and $K_-=-1$ so that the perturbation is supported on sites $0$ and $1$. Then
\begin{align*}
&
{\Mm}^z(1,-1)
\;=\;
\left[
\one
\;+\;
\imath\,\nu^z
\begin{pmatrix}
V(1) & z^{-2}V(1) \\ -z^2V(1) & -V(1) 
\end{pmatrix}
\right]
\left[
\one
\;+\;
\imath\,\nu^z
\begin{pmatrix}
V(0) &  V(0) \\ - V(0) & -V(0) 
\end{pmatrix}
\right]
\;.
\end{align*}
which allows to read off

\begin{align*}
M^z_-
&
\;=\;
\one -\imath\,\nu^z(V(0)+V(1)) +(\nu^z)^2(z^2-1)V(0)V(1)
\\
&
\;=\;
\one -\imath\,\nu^z(V(0)+V(1)-zV(0)V(1))
\\
&
\;=\;
\one -\tfrac{z}{z+1}\,V(0)V(1)-\imath\,\nu^z(V(0)+V(1)-V(0)V(1))
\;,
\end{align*}
where the identity $\nu^z(z^2-1)=\imath z$ was used. Recalling that $M_-^z$ is invertible for $|z|=1$ (see Lemma \ref{lem-Minvertible}), the transmission coefficient in this case is thus
$$
T^z_-
\;=\;
\left(
\one -\imath\,\nu^z(V(0)+V(1)-zV(0)V(1))
\right)^{-1}
\;.
$$
In the notations of Section~\ref{sec-BandEdge}, one can hence read off that  $G^z=\one -\tfrac{z}{z+1}\,V(0)V(1)$ and $F=-\imath (V(0)+V(1)-V(0)V(1))$.  It is now readily possible to construct examples with singular $F$ and hence non-generic band edge asymptotics.

\vspace{.3cm}

\noindent {\bf Acknowledgements:} "Research supported by Conacyt, FORDECYT-PRONACES-29825-2020". The work of M.~B. and G.~F. was supported by the proyects PAPIIT-DGAPA-UNAM IN108818 and PAPIIT-DGAPA-UNAM IN101621.  That of H.~S.-B. by PAPIIT-UNAM IN105718, CONACYT Ciencia B\'asica 283531 and the DFG.  Research partially supported by project PAPIIT-DGAPA UNAM IN103918


\end{document}